\documentclass[letter, amsfonts, amssymb, amsmath, reprint, showkeys, nofootinbib, twoside, superscriptaddress]{revtex4-1}
\usepackage{color}
\usepackage{amsmath}
\usepackage{pifont}   
\usepackage{verbatim}       
\usepackage{acronym} 

\usepackage{amsmath}
\usepackage{amsfonts}
\usepackage{amssymb}
\usepackage{mathrsfs} 
\usepackage{graphicx}
\usepackage{multirow} 
\usepackage{slashed} 
\usepackage{array}
\usepackage{bbm}

\usepackage{graphicx}
\usepackage{epsf}
\usepackage{bm}
\usepackage{booktabs}
\usepackage{array}
\usepackage{caption}
\usepackage{subcaption}
\usepackage[utf8]{inputenc}

\allowdisplaybreaks

\usepackage[pdftex, pdftitle={Article}, pdfauthor={Author}]{hyperref} % For hyperlinks in the PDF

\begin{document}

\title{Observability of the Higgs boson decay to a photon and a dark photon}

\author{Hugues Beauchesne}
  \email[Email address: ]{beauchesneh@phys.ncts.ntu.edu.tw}
  \affiliation{Physics Division, National Center for Theoretical Sciences, Taipei 10617, Taiwan}

\author{Cheng-Wei Chiang}
  \email[Email address: ]{chengwei@phys.ntu.edu.tw}
  \affiliation{Department of Physics and Center for Theoretical Physics, National Taiwan University, Taipei 10617, Taiwan}
  \affiliation{Physics Division, National Center for Theoretical Sciences, Taipei 10617, Taiwan}

\begin{abstract}
Many collider searches have attempted to detect the Higgs boson decaying to a photon and an invisible massless dark photon. For the branching ratio to this channel to be realistically observable at the LHC, there must exist new mediators that interact with both the standard model and the dark photon. In this paper, we study experimental and theoretical constraints on an extensive set of mediator models. We show that these constraints limit the Higgs branching ratio to a photon and a dark photon to be far smaller than the current sensitivity of collider searches.
\end{abstract}

\maketitle

%%%%%%%%%%%%%%%%%%%%%%%%%%%%%%%%%%%%%%%%%%%%%%%%%%
\section{Introduction}\label{Sec:Intro}
%%%%%%%%%%%%%%%%%%%%%%%%%%%%%%%%%%%%%%%%%%%%%%%%%%

The dark photon is a hypothetical Abelian gauge boson that can mix with the photon \cite{Holdom:1985ag} and has been extensively studied both experimentally and theoretically. It could, for example, act as a portal to a dark sector or introduce dark matter self-interactions that can potentially solve the small-scale structure problems \cite{Spergel:1999mh} and the XENON1T anomaly \cite{XENON:2020rca,Chiang:2020hgb}.  Depending on its decoupling temperature during the evolution of the Universe, the dark photon can affect the big bang nucleosynthesis by altering the effective number of thermally excited neutrino degrees of freedom \cite{Dobrescu:2004wz,Fradette:2014sza}.  The dark photon may also modify the stellar energy transport mechanism and thus the cooling of, for example, neutron stars \cite{Lu:2021uec}.

A potential discovery channel of the dark photon that has received considerable attention is the decay of the Higgs boson to a photon and a massless dark photon. The interest in this channel stems in large part from early phenomenological studies \cite{Gabrielli:2014oya, Biswas:2015sha, Biswas:2016jsh, Biswas:2017lyg, Biswas:2017anm}, which indicated that a corresponding branching ratio as large as $5\%$ was at the time compatible with experimental constraints \cite{Gabrielli:2014oya}. This has motivated several collider searches for the Higgs boson decaying to a photon and a dark photon \cite{CMS:2019ajt, CMS:2020krr, ATLAS:2021pdg, ATLAS:2022xlo}. Presently, the most precise collider bound on this branching ratio comes from Ref.~\cite{ATLAS:2021pdg} by ATLAS, which uses 139~fb$^{-1}$ of integrated luminosity at a center-of-mass energy of 13~TeV. This study has obtained an upper limit of $1.8\%$ at 95\% confidence level (CL).

For the Higgs boson to be able to decay to a photon and a dark photon, there must exist some mechanism that allows interactions between the dark photon and standard model (SM) particles. In principle, SM particles themselves could interact at tree level with the dark photon. For example, this could be because of kinematic mixing between the weak hypercharge and the gauge boson of a new $U(1)'$ gauge group. The problem with this scenario, however, is that the branching ratio of the Higgs boson to photons is small and that interactions between SM particles and a new light gauge boson are constrained to be very small (see, e.g., \cite{Bjorken:1988as, Curtin:2014cca, Chang:2016ntp, Fox:2018ldq, Parker:2018vye, Pan:2018dmu}) or can sometimes outright be rotated away \cite{Fabbrichesi:2020wbt}. The branching ratio of the Higgs boson to a photon and a dark photon would therefore be far smaller than what could realistically be observed at the LHC. A potentially discoverable branching ratio of the Higgs boson to a photon and a dark photon then requires the introduction of new particles that can mediate interactions between the SM sector and the dark photon.  Therefore, an observation of this channel would not only verify the existence of the dark photon, but also provide indirect evidence of more new particles.

In this paper, we study constraints on mediators that enable the Higgs boson to decay to a photon and a massless invisible dark photon. Crucially, we demonstrate that these constraints restrict this branching ratio to values considerably lower than current collider limits. To do this, we consider constraints from the Higgs signal strengths, electroweak precision tests, the electric dipole moment (EDM) of the electron, and unitarity. This paper is an extended and more detailed version of Ref.~\cite{PhysRevLett.130.141801}.

Despite previous claims, it is not practically possible to obtain a bound on the branching ratio of the Higgs boson to a photon and a dark photon that is completely model-independent. Certain crucial constraints like the Higgs signal strengths are too elaborate and can be affected by too many factors for a bound to be strictly universal. To counteract this, we will consider an extensive and representative set of mediator models. They constitute the complete set of possible models that respect some very minimal assumptions, which we will present below. These models are, of course, ultimately only benchmarks, but they will clearly illustrate the reasons why, given the above-mentioned constraints, building models that lead to a large branching ratio of the Higgs boson to a photon and a dark photon would be extremely challenging.

We find the following results. In all models considered, the branching ratio of the Higgs boson to a photon and a dark photon is constrained to be below $\sim 0.4\%$. Furthermore, for many models, this number is only technically allowed because of the absence of collider searches for certain hard-to-miss signatures and would be considerably lower given the existence of such searches. For certain mediators, the constraints are even considerably stronger than 0.4\%.\footnote{Ref.~\cite{Biswas:2022tcw} contains conclusions similar to ours, though the authors did not analyze constraints as extensively and as such did not obtain limits quite as strong.}

This paper is organized as follows. In Sec.~\ref{Sec:Assumptions}, we introduce our assumptions and the models they allow. Sections~\ref{Sec:FermionMediators} and~\ref{Sec:ScalarMediators} present the constraints and results for the fermion and scalar mediators, respectively. Concluding remarks are presented in Sec.~\ref{Sec:Conclusion}, including a discussion on the effects of relaxing our assumptions.

%%%%%%%%%%%%%%%%%%%%%%%%%%%%%%%%%%%%%%%%%%%%%%%%%%%%%%%%%%%%
\section{Assumptions and models}\label{Sec:Assumptions}
%%%%%%%%%%%%%%%%%%%%%%%%%%%%%%%%%%%%%%%%%%%%%%%%%%%%%%%%%%%%

We begin by presenting our assumptions on how the dark photon and the SM can communicate. Of course, there would be ways to violate them and we discuss the implications of this in the conclusion. The models they admit are presented here in a schematic way. The dark photon is referred to as $A'$, the photon as $A$, and the Higgs boson as $h$.

Assume a new $U(1)'$ gauge group whose gauge boson is $A'$ and that all SM particles are neutral under this group. Assume a set of mediators charged under both SM gauge groups and the $U(1)'$. Then, the conditions that we require our mediator models to satisfy are as follows:
\begin{enumerate}
  \item The Lagrangian is renormalizable and preserves all the gauge symmetries.
  \item The Higgs decay to $AA'$ can occur at one loop.
  \item The mediators are neutral under QCD.
  \item The mediators are either complex scalars or vector-like fermions.
  \item There are no more than two new fields.
  \item There are no mediators that have a nonzero expectation value or mix with SM fields.
\end{enumerate}
Several comments are in order: 
\begin{itemize}
  \item Assumption 1 is a standard requirement of beyond the standard model (BSM) physics. In addition, a non-renormalizable Lagrangian would allow for tree-level decay of the Higgs boson to $AA'$, which would render the analysis trivial but leave unclear whether a reasonable UV completion  is possible.
  \item Assumption 2 is required such that $\text{BR}(h \to A A')$ be sufficiently large to observe. If this decay were to take place at an even higher loop level, the branching ratio would generally simply be too small. In principle, a large $\text{BR}(h \to A A')$ could be obtained without this assumption being satisfied in the presence of a new non-perturbative sector, but the complicated nature of this would make the analysis less definite and we do not consider it further. This decay is forbidden by gauge invariance to take place at tree level for a renormalizable Lagrangian. 
  \item Assumption 3 is made for two reasons. First, a mediator charged under QCD would be forced to have a mass at the TeV scale, which would make it difficult to obtain a large $\text{BR}(h \to A A')$. Second, even if a large $\text{BR}(h \to A A')$ could somehow be obtained, it would unavoidably imply a large modification of the Higgs interactions with the gluons. This would be in tension with experimental measurements, especially considering the gluon-fusion cross section is known at $\mathcal{O}(10\%)$ precision.
  \item Assumption 4 is made since there are not many well-motivated BSM models containing mediators with spin higher than $1/2$ that lead to a sizable $\text{BR}(h \to A A')$.
  \item Assumptions 4-6 are not, in principle, mandatory, but are introduced to keep the number of possible models at a manageable level.
\end{itemize}

The most important consequence of these assumptions is that they require the Lagrangian to contain a term coupling the Higgs doublet to mediators at tree level. There are only five generic forms such a term can take while respecting our assumptions. Each class of mediator models then corresponds to a different form of the Higgs coupling to mediators. The classes consist of a single fermion class and four scalar classes. The possibilities for these interaction terms are in a schematic form:
\begin{equation}\label{eq:ModelList}
  \begin{aligned}
    &\text{Fermion:}\\
    & \quad \overline{\psi}_1 (A_L P_L + A_R P_R) \psi_2 H + \text{H.c.},  \hspace{-10cm}\\
    &\text{Scalar:}\\
    & \quad \text{I:}   && \mu\phi_1^\dagger \phi_2 H + \text{H.c.}, &
                    & \quad \text{II:}  && \lambda H^\dagger H \phi^\dagger \phi,\\
    & \quad \text{III:} && \lambda H^\dagger H \phi_1^\dagger \phi_2 + \text{H.c.}, &
                    & \quad \text{IV:}  && \lambda H H \phi_1^\dagger \phi_2 + \text{H.c.},
  \end{aligned}
\end{equation}
where $H$ is the Higgs doublet, and the different $\psi$'s are vector-like fermions and $\phi$ scalars. All indices are suppressed and the details will be explained in Secs.~\ref{Sec:FermionMediators} and \ref{Sec:ScalarMediators}. For each class, the fields can take different quantum numbers and this is why we say that they are classes of models. These models could, of course, be combined, but we will always consider one model at a time for the sake of definiteness and manageability. In the rest of this paper, we will elaborate on each model and study how large of a $\text{BR}(h \to A A')$ they can potentially lead to.

%%%%%%%%%%%%%%%%%%%%%%%%%%%%%%%%%%%%%%%%%%%%%%
\section{Fermion mediators}\label{Sec:FermionMediators}
%%%%%%%%%%%%%%%%%%%%%%%%%%%%%%%%%%%%%%%%%%%%%%

We present in this section the only class of fermion mediator models that respect the assumptions listed in Sec.~\ref{Sec:Assumptions}. The different constraints and the allowed $\text{BR}(h \to A A')$ are also discussed.

\subsection{Field content, Lagrangian, and mass eigenstates}\label{sSec:CaseILagrangian}
We begin by introducing the relevant fields and Lagrangian. Consider a vector-like fermion $\psi_1$ that transforms under a representation of $SU(2)_L$ of dimension $p = n \pm 1$, has a weak hypercharge of $Y^p = Y^n + 1/2$ and a charge $Q'$ under $U(1)'$. Consider another vector-like fermion $\psi_2$ that transforms under a representation of $SU(2)_L$ of dimension $n$, has a weak hypercharge of $Y^n$, and a charge $Q'$ under $U(1)'$. The Lagrangian that determines the masses of the fermions is
\begin{equation}\label{eq:CaseILagrangian}
  \begin{aligned}
  \mathcal{L}_m = &-\left[\sum_{a,b,c}\hat{d}^{pn}_{abc} \overline{\psi}_1^a (A_L P_L + A_R P_R) \psi_2^b H^c + \text{H.c.}\right]\\
  &- \mu_1 \overline{\psi}_1 \psi_1 - \mu_2 \overline{\psi}_2 \psi_2.
  \end{aligned}
\end{equation}
where $a$, $b$, and $c$ are $SU(2)_L$ indices and are summed from 1 (corresponding to the highest weight state) to the size of the corresponding multiplet.  We will write explicitly $SU(2)_L$ indices and sums when they are non-trivial. The $SU(2)_L$ tensor $\hat{d}^{pn}_{abc}$ is uniquely fixed by gauge invariance and given by the Clebsch-Gordan coefficient
\begin{equation}\label{eq:dnabc1}
  \hat{d}^{pn}_{abc} = C^{JM}_{j_1 m_1 j_2 m_2} = \langle j_1 j_2 m_1 m_2 |J M \rangle,
\end{equation}
where
\begin{equation}\label{eq:dnabc2}
  \begin{aligned}
    J &= \frac{p -1}{2},       & j_1 &= \frac{n - 1}{2},      & j_2 &= \frac{1}{2}, \\ 
    M &= \frac{p + 1 - 2a}{2}, & m_1 &= \frac{n + 1 - 2b}{2}, & m_2 &=\frac{3 - 2c}{2}.
  \end{aligned}
\end{equation}
We will assume throughout this paper that, as is most commonly the case, the phase convention of the Clebsch-Gordan coefficients is chosen such that they are always real. We will further assume that the Clebsch-Gordan coefficients that correspond to non-physical spin combinations are zero. These two assumptions will lighten the notation. The parameters $A_L$ and $A_R$ can be complex, and it is easy to verify that there is a complex phase that cannot generally be reabsorbed via field redefinition.  Such a physical complex phase will lead to CP-violating interactions.  Once the Higgs field obtains a vacuum expectation value (VEV), the Lagrangian $\mathcal{L}_m$ will contain the mass terms
\begin{equation}\label{eq:CaseILagrangianmassI}
  \begin{aligned}
  \mathcal{L}_m \supset &-\Biggl[\sum_{a,b}\frac{A_L v}{\sqrt{2}} \hat{d}^{pn}_{ab2} \overline{\psi}^a_1 P_L \psi_2^b + \sum_{a,b}\frac{A_R^\ast v}{\sqrt{2}} \hat{d}^{pn}_{ba2} \overline{\psi}_2^a P_L \psi_1^b \\
  &  \hspace{0.5cm} + \mu_1 \overline{\psi}_1 P_L \psi_1 + \mu_2 \overline{\psi}_2 P_L \psi_2 + \text{H.c.}\Biggr],
  \end{aligned}
\end{equation}
where $v \approx 246$~GeV is the VEV of the Higgs field. Introduce the convenient notation
\begin{equation}\label{eq:CaseINotationI}
  \hat{\psi} = \begin{pmatrix} \psi_1 \\ \psi_2 \end{pmatrix}
\end{equation}
and
\begin{equation}\label{eq:CaseINotationII}
  \quad d^{pn}_{ab} = \begin{cases} \hat{d}^{pn}_{a(b - p)2}, & \text{if $a \in [1, p]$ and $b \in [p + 1, p + n]$},\\ 0, & \text{otherwise.}\end{cases}\\
\end{equation}
The mass Lagrangian can be written more succinctly as
\begin{equation}\label{eq:CaseILagrangianmassII}
  \mathcal{L}_m \supset -\sum_{a,b} M_{ab} \overline{\hat{\psi}}^a P_L \hat{\psi}^b + \text{H.c.},
\end{equation}
where the mass matrix is
\begin{equation}\label{eq:CaseIMassMatrix}
  M = \begin{pmatrix} \mu_1 \mathbbm{1}_{p\times p} & 0_{p\times n} \\ 0_{n\times p} & \mu_2 \mathbbm{1}_{n\times n} \end{pmatrix}  + \frac{A_L v}{\sqrt{2}} d^{pn}+ \frac{A_R^\ast v}{\sqrt{2}} d^{pnT}.
\end{equation}
The mass matrix can then be diagonalized by introducing the fields $\tilde{\psi}^a$ via
\begin{equation}\label{eq:CaseImasseigenstates}
 P_L \hat{\psi} = R_L P_L \tilde{\psi}, \quad P_R \hat{\psi} = R_R P_R \tilde{\psi},
\end{equation}
where $R_L$ and $R_R$ are unitary matrices that diagonalize $M^\dagger M$ and $MM^\dagger$, respectively. Of course, these matrices only mix particles of identical electric charges and all entries that correspond to mixing of particles of different charges are zero. The fields $\tilde{\psi}^a$ are the mass eigenstates of mass $m_a$, and there are $p + n$ of them.

Finally, the interactions of the Higgs boson with the mass eigenstates $\tilde{\psi}^a$ are controlled by the terms
\begin{equation}\label{eq:CaseIHiggsInteractionsI}
  \mathcal{L}_m \supset -\sum_{a,b} \Omega_{ab} h \overline{\tilde{\psi}}^a P_L \tilde{\psi}^b + \text{H.c.},
\end{equation}
where $\Omega$ is generally neither Hermitian nor diagonal and given by
\begin{equation}\label{eq:CaseIHiggsInteractionsII}
   \quad \Omega = \frac{A_L}{\sqrt{2}} R_R^\dagger d^{pn} R_L + \frac{A_R^\ast}{\sqrt{2}} R_R^\dagger d^{pnT} R_L.
\end{equation}

\subsection{Gauge interactions}\label{sSec:CaseIInteractions}
We now discuss all relevant gauge interactions of the mass eigenstates $\tilde{\psi}^a$. The interactions of the $A/A'$ with $\tilde{\psi}^a$ are controlled by
\begin{equation}\label{eq:CaseIPhotonInteractionsI}
  \mathcal{L}_g \supset -e A_\mu \overline{\tilde{\psi}}\gamma^\mu \tilde{Q}\tilde{\psi} - Q'e' A_\mu' \overline{\tilde{\psi}}\gamma^\mu \tilde{\psi}, 
\end{equation}
where $e'$ is the gauge coupling constant of $U(1)'$ and $\tilde{Q}$ is the diagonal charge matrix
\begin{equation}\label{eq:CaseIPhotonInteractionsII}
  \tilde{Q} = R_L^\dagger \hat{Q} R_L = R_R^\dagger \hat{Q} R_R,
\end{equation}
as QED is a vector-like interaction,
where
\begin{equation}\label{eq:CaseIPhotonInteractionsIII}
  \hat{Q} = \begin{pmatrix} Y^p + T_3^p & 0_{p\times n} \\ 0_{n\times p} & Y^n + T_3^n \end{pmatrix},
\end{equation}
with $(T_3^p)_{ab} = (p + 1 - 2a)\delta_{ab}/2$ and similarly for $T_3^n$.
In practice, $\tilde{Q}$ is identical to $\hat{Q}$ except for a potential reordering of the diagonal elements.

The interactions between the $Z$ boson and $\tilde{\psi}^a$ is controlled by the terms
\begin{equation}\label{eq:CaseIZInteractionsI}
  \mathcal{L}_g \supset -\sqrt{g^2 + {g'}^2} Z_\mu \overline{\tilde{\psi}}\gamma^\mu (B_L P_L + B_R P_R) \tilde{\psi},
\end{equation}
where $B_L$ and $B_R$ are, in general, non-diagonal but Hermitian and given by
\begin{equation}\label{eq:CaseIZInteractionsII}
  \begin{aligned}
    B_L &= R_L^\dagger\begin{pmatrix} -s_W^2 Y^p + c_W^2 T_3^p & 0_{p\times n} \\ 0_{n\times p} & -s_W^2 Y^n + c_W^2 T_3^n \end{pmatrix} R_L,\\
    B_R &= R_R^\dagger\begin{pmatrix} -s_W^2 Y^p + c_W^2 T_3^p & 0_{p\times n} \\ 0_{n\times p} & -s_W^2 Y^n + c_W^2 T_3^n \end{pmatrix} R_R,
  \end{aligned}
\end{equation}
with $g$ ($g'$) denoting the $SU(2)_L$ ($U(1)_Y$) gauge coupling constant and $s_W$ ($c_W$) the sine (cosine) of the weak angle.

The interactions between the $W$ boson and $\tilde{\psi}^a$ is controlled by the terms
\begin{equation}\label{eq:CaseIWInteractionsI}
  \mathcal{L}_g \supset -\frac{g}{\sqrt{2}} \overline{\tilde{\psi}}\gamma^\mu\left(\hat{A}_L P_L W^+_\mu + \hat{A}_R P_R W^+_\mu\right)\tilde{\psi} + \text{H.c.},
\end{equation}
where
\begin{equation}\label{eq:CaseIWInteractionsII}
  \hat{A}_L = R_L^\dagger\begin{pmatrix} T_+^p & 0_{p\times n} \\ 0_{n\times p} & T_+^n \end{pmatrix} R_L, \;
  \hat{A}_R = R_R^\dagger\begin{pmatrix} T_+^p & 0_{p\times n} \\ 0_{n\times p} & T_+^n \end{pmatrix} R_R,
\end{equation}
with $(T_+^p)_{ab} = \sqrt{a(p - a)}\delta_{a,b-1}$ and similarly for $T_+^n$.

\subsection{Relevant Higgs decays}\label{sSec:CaseIHiggsdecay}
\begin{figure}[t!]
%\begin{center}
 \captionsetup[subfigure]{justification=centerlast}
 \begin{subfigure}{0.4\textwidth}
    \centering
    \caption{}
    \includegraphics[width=1\textwidth]{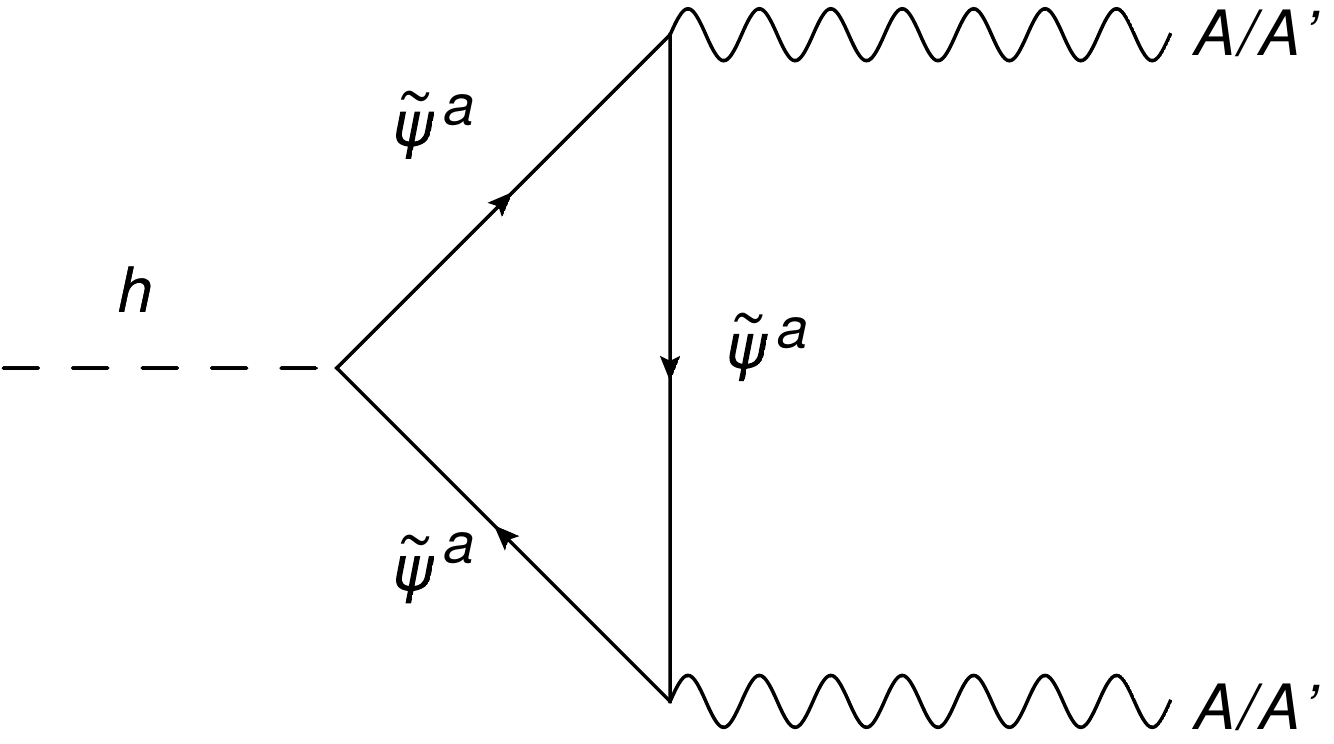}
    \label{fig:CaseIHiggsDecay1}
  \end{subfigure}\qquad
  \captionsetup[subfigure]{justification=centerlast}
  \begin{subfigure}{0.4\textwidth}
    \centering
    \caption{}
    \includegraphics[width=1\textwidth]{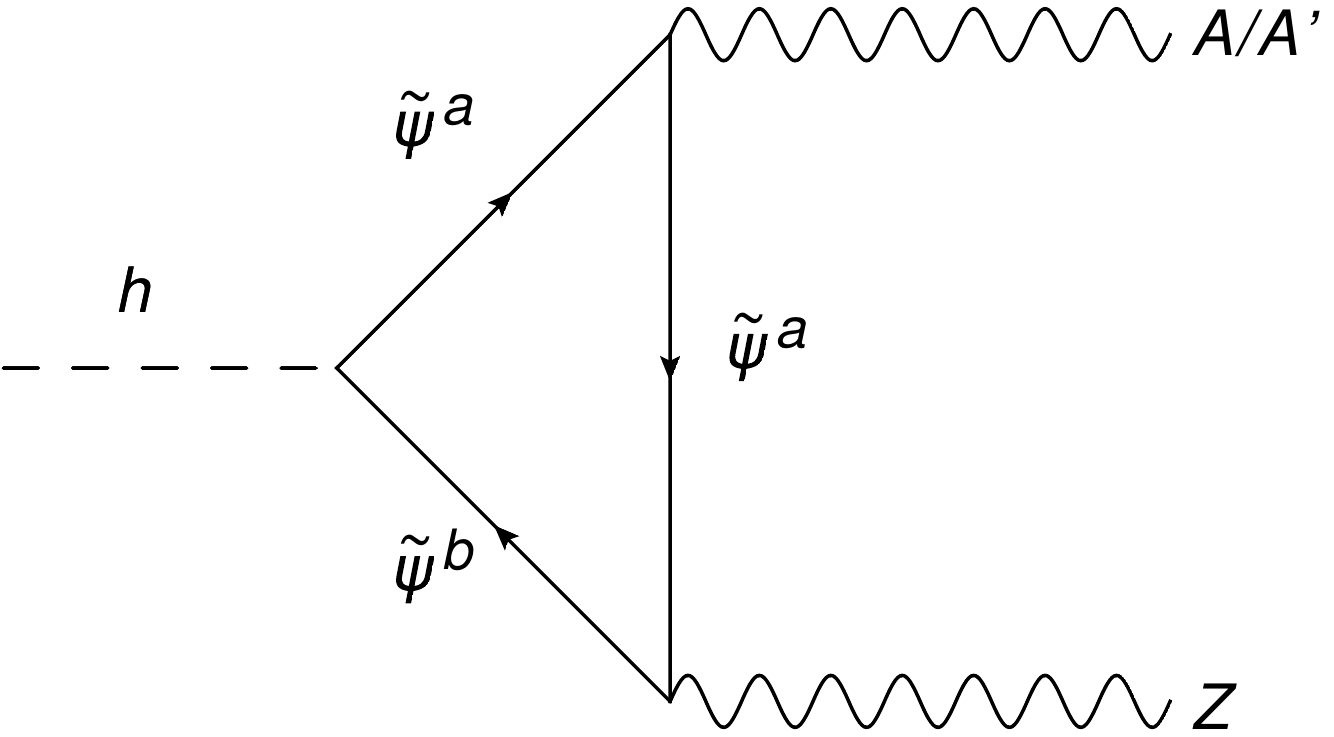}
    \label{fig:CaseIHiggsDecay2}
  \end{subfigure}
\caption{(a) Higgs decay to two $A/A'$. (b) Higgs decay to a $Z$ boson and $A/A'$. Diagrams with the flow of the mediators inverted also exist.
}
\label{fig:CaseIHiggsDecay}
%\end{center}
\end{figure}

The interactions of the mediators $\tilde{\psi}^a$ lead to contributions to the amplitude of the Higgs decay to $AA'$, but also to those of the experimentally constrained decays to $AA$ and $A'A'$. The relevant diagrams are shown in Fig.~\ref{fig:CaseIHiggsDecay1}. Irrespective of the mediators, gauge invariance forces the amplitudes to take the forms
\begin{equation}\label{eq:CaseIHiggsAAdecayI}
  \begin{aligned}
    M^{h\to AA} = & S^{h\to AA} \left(p_1\cdot p_2 g_{\mu\nu} - p_{1\mu} p_{2\nu}\right)\epsilon^\nu_{p_1}\epsilon^\mu_{p_2}\\
                & + i\tilde{S}^{h\to AA}\epsilon_{\mu\nu\alpha\beta}p_1^\alpha p_2^\beta\epsilon^\nu_{p_1}\epsilon^\mu_{p_2},\\
    M^{h\to AA'} = & S^{h\to AA'} \left(p_1\cdot p_2 g_{\mu\nu} - p_{1\mu} p_{2\nu}\right)\epsilon^\nu_{p_1}\epsilon^\mu_{p_2}\\
                & + i\tilde{S}^{h\to AA'}\epsilon_{\mu\nu\alpha\beta}p_1^\alpha p_2^\beta\epsilon^\nu_{p_1}\epsilon^\mu_{p_2},\\
    M^{h\to A'A'} = & S^{h\to A'A'} \left(p_1\cdot p_2 g_{\mu\nu} - p_{1\mu} p_{2\nu}\right)\epsilon^\nu_{p_1}\epsilon^\mu_{p_2}\\
                & + i\tilde{S}^{h\to A'A'}\epsilon_{\mu\nu\alpha\beta}p_1^\alpha p_2^\beta\epsilon^\nu_{p_1}\epsilon^\mu_{p_2},\\
  \end{aligned}
\end{equation}
where $p_1$ and $p_2$ are the momenta of the two gauge bosons. The $S$ coefficients are CP conserving, while the $\tilde S$ coefficients are CP violating. For the fermion mediators, the coefficients are given at one loop by
\begin{equation}\label{eq:CaseIHiggsAAdecayII}
  \begin{split}
    S^{h\to AA  } &= e^2    \sum_a \text{Re}(\Omega_{aa})\tilde{Q}_{aa}^2  S_a + S^{h\to AA}_\text{SM}, \\
    S^{h\to AA' } &= e e'   \sum_a \text{Re}(\Omega_{aa})\tilde{Q}_{aa} Q' S_a,                         \\
    S^{h\to A'A'} &= {e'}^2 \sum_a \text{Re}(\Omega_{aa}){Q'}^2            S_a,                         \\
    \tilde{S}^{h\to AA  } &= e^2    \sum_a \text{Im}(\Omega_{aa})\tilde{Q}_{aa}^2  \tilde{S}_a + \tilde{S}^{h\to AA}_\text{SM},\\
    \tilde{S}^{h\to AA' } &= e e'   \sum_a \text{Im}(\Omega_{aa})\tilde{Q}_{aa} Q' \tilde{S}_a,\\
    \tilde{S}^{h\to A'A'} &= {e'}^2 \sum_a \text{Im}(\Omega_{aa}){Q'}^2            \tilde{S}_a,
  \end{split}
\end{equation}
where $S^{h\to AA}_\text{SM}\approx 3.3 \times 10^{-5}$~GeV$^{-1}$ and $\tilde{S}^{h\to AA}_\text{SM} \approx 0$~GeV$^{-1}$ are the SM contributions to their respective coefficients and
\begin{equation}\label{eq:CaseIHiggsAAdecayIII}
  \begin{aligned}
    S_a         &= \frac{-m_a}{2\pi^2 m_h^2}\left[ 2 + (4 m_a^2 - m_h^2)C_0(0, 0, m_h^2; m_a, m_a, m_a) \right],\\
    \tilde{S}_a &= -i\frac{m_a}{2\pi^2}C_0(0, 0, m_h^2; m_a, m_a, m_a),
  \end{aligned}
\end{equation}
with $C_0(s_1, s_{12}, s_2; m_0, m_1, m_2)$ being the scalar three-point Passarino-Veltman function \cite{Passarino:1978jh}.\footnote{All loop calculations in this paper are performed with the help of Package-X \cite{Patel:2015tea}.} The decay widths are then given by
\begin{equation}\label{eq:CaseIDecayWidths}
  \begin{gathered}
    \Gamma^{h\to AA  } = \frac{|S^{h\to AA  }|^2 + |\tilde{S}^{h\to AA  }|^2}{64\pi}m_h^3, \\
    \Gamma^{h\to AA' } = \frac{|S^{h\to AA' }|^2 + |\tilde{S}^{h\to AA' }|^2}{32\pi}m_h^3, \\
    \Gamma^{h\to A'A'} = \frac{|S^{h\to A'A'}|^2 + |\tilde{S}^{h\to A'A'}|^2}{64\pi}m_h^3,
  \end{gathered}
\end{equation}
where a symmetry factor of $1/2$ has been included for the $AA$ and $A'A'$ final states. The decay of the Higgs boson to a $Z$ boson and either $A$ or $A'$ is shown in Fig.~\ref{fig:CaseIHiggsDecay2} and has an amplitude similar in form to Eq.~\eqref{eq:CaseIHiggsAAdecayII}, albeit with much more complicated coefficients $S^{h\to ZA}$, $\tilde{S}^{h\to ZA}$, $S^{h\to ZA'}$, and $\tilde{S}^{h\to ZA'}$ due to the possibility of two different mediators running in the loop.

There are several crucial results of this section that need to be discussed. First, the presence of the Levi-Civita tensor in the amplitudes is due to the $\gamma^5$ Dirac matrix in the Higgs to two mediators vertex. There will be no such term for the scalar cases.

Second, the amplitudes of Eq.~\eqref{eq:CaseIHiggsAAdecayI} are all highly correlated. Because of this, a large $\text{BR}(h \to A A')$ will generally lead to either a large branching ratio of the Higgs boson to invisible particles, a large modification of the coupling of the Higgs to photons, or both. The Higgs signal strengths will therefore impose strong constraints on $\text{BR}(h \to A A')$.

Third, the decay width into two photons will in general contain terms that come from the interference between the SM amplitudes and the mediator amplitudes. When present, these cross terms typically dominate the modifications of the decay width. The presence of cross terms could, in principle, be avoided in two ways. The first way would be for the mediators to only provide a purely imaginary contribution to $S^{h\to AA}$. Since the SM contribution to $S^{h\to AA}$ is almost purely real, there would be essentially no interference. However, this turns out to be impossible for the fermion mediators. As seen in Eq.~\eqref{eq:CaseIHiggsAAdecayII}, all constants appearing in the $S^{h\to AA}$ are real. In addition, the kinematic function $S_a$ must be purely real, as bounds from the Large Electron-Positron collider (LEP) prevent charged mediators from being sufficiently light to be able to ``cut'' the diagram \cite{LEP1, LEP2}. The second way to avoid interference terms would be for the mediators to only contribute to $\tilde{S}^{h\to AA}$. This can indeed be done for fermion mediators and the signal strength constraints can therefore mostly be evaded. However, a large $\text{BR}(h \to A A')$ will in this case lead to a large EDM for the electron. The limits on the EDM will force the complex phase to be small and will close this loophole.

Fourth, it is possible at this point to perform a naive estimate of the upper limit on $\text{BR}(h \to AA')$ allowed by the Higgs signal strengths. A sufficiently large $\text{BR}(h \to AA')$ will imply large Yukawa couplings. This will generally split the masses of the different $\tilde{\psi}^a$ and lead to a single mediator dominating the amplitude. Assume as justified above that $\text{Im}(\Omega_{ii}) = \text{Im}(S_a) = 0$.  Call $\Delta \text{BR}(h \to AA)$ the deviation of $\text{BR}(h \to AA)$ from its SM value and assume that it is small. Then, the following approximate relation holds:
\begin{equation}\label{eq:CaseIAmplitudeRelation}
  \begin{aligned}
  & \text{BR}(h \to AA') \approx\\
  & \hspace{0.5 cm} \sqrt{\text{BR}(h\to A'A') \text{BR}(h\to AA)} \left|\frac{\Delta \text{BR}(h \to AA)}{\text{BR}(h\to AA)}\right|.
  \end{aligned}
\end{equation}
The branching ratio $\text{BR}(h \to AA)$ is about $0.23\%$ and can at most deviate by $\mathcal{O}(25\%)$ from this value. The branching ratio of the Higgs to invisible particles $\text{BR}(h\to A'A')$ is at most $\mathcal{O}(10\%)$. This means that $\text{BR}(h \to AA')$ is at most $\mathcal{O}(0.4\%)$. We will see that this approximation holds well, though other constraints will often force $\text{BR}(h \to AA')$ to be even smaller. Of course, Eq.~\eqref{eq:CaseIAmplitudeRelation} assumes that the amplitude is dominated by a single mediator, which may not be a good approximation in more complicated models. Deviations will be observed when multiple mediators contribute comparably to $\text{BR}(h \to AA')$, but the limits will remain of the same order of magnitude barring large fine-tuning or elaborate model building.

\subsection{Higgs signal strengths}\label{sSec:CaseIHiggsSS}
The constraints associated with the Higgs signal strengths are taken into account by using the $\kappa$ formalism \cite{Heinemeyer:2013tqa}. Assume a production mechanism $i$ with cross section $\sigma_i$ or a decay process $i$ with width $\Gamma_i$. The parameter $\kappa_i$ is defined such that 
\begin{equation}\label{eq:Defkappa}
  \kappa_i^2 = \frac{\sigma_i}{\sigma_i^{\text{SM}}} \quad \text{or} \quad \kappa_i^2 = \frac{\Gamma_i}{\Gamma_i^{\text{SM}}},
\end{equation}
where $\sigma_i^{\text{SM}}$ and $\Gamma_i^{\text{SM}}$ are the corresponding SM quantities. The only two Higgs couplings that are affected at leading order are those associated with $AA$ and $AZ$. The corresponding $\kappa$'s are
\begin{equation}\label{eq:CaseIKappaI}
  \begin{aligned}
  \kappa_{AA}^2 = \frac{|S^{h\to AA}|^2 + |\tilde{S}^{h\to AA}|^2}{|S^{h\to AA}_\text{SM}|^2 + |\tilde{S}^{h\to AA}_\text{SM}|^2},\\
  \kappa_{ZA}^2 = \frac{|S^{h\to ZA}|^2 + |\tilde{S}^{h\to ZA}|^2}{|S^{h\to ZA}_\text{SM}|^2 + |\tilde{S}^{h\to ZA}_\text{SM}|^2}.
  \end{aligned}
\end{equation}
The invisible ($A'A'$) and semi-invisible ($A'A$ and $A'Z$) decays of the Higgs boson are taken into account by properly rescaling the signal strengths. The resulting global reduction of the Higgs signal strengths renders constraints from searches for the Higgs decay to invisible particles mostly superfluous and we do not impose them \cite{ATLAS:2023tkt}. Constraints are then applied by using the Higgs signal strength measurements of Ref.~\cite{CMS-PAS-HIG-19-005} by CMS, which uses $137$ fb$^{-1}$ of integrated luminosity at 13~TeV center-of-mass energy, and Ref.~\cite{ATLAS-CONF-2021-053} by ATLAS, which uses $139$ fb$^{-1}$ of integrated luminosity at also 13~TeV. These studies conveniently provide all the information necessary (measurements, uncertainties, and correlations) to produce our own $\chi^2$ fit. The two searches are assumed to be uncorrelated. As we will be interested in two-dimensional scans, a point of parameter space will be considered excluded at 95\% CL if its $\chi^2$ satisfies $\chi^2 - \chi^2_{\text{min}} > 5.99$, where $\chi^2_{\text{min}}$ is the best fit of the model.\footnote{Technically speaking, it is possible for the mediator contributions to $S^{h\to AA}$ to be about $-2S^{h\to AA}_\text{SM}$. This would mostly avoid the bounds from the Higgs signal strengths and could in principle lead to a larger $\text{BR}(h \to AA')$. It however requires very exotic circumstances and a very large amount of fine-tuning. As such, we will not consider such cases.}

\subsection{Electron EDM}\label{sSec:CaseIElectronEDM}
\begin{figure}[t!]
%\begin{center}
 \captionsetup[subfigure]{justification=centerlast}
 \begin{subfigure}{0.3\textwidth}
    \centering
    \caption{}
    \includegraphics[width=1\textwidth]{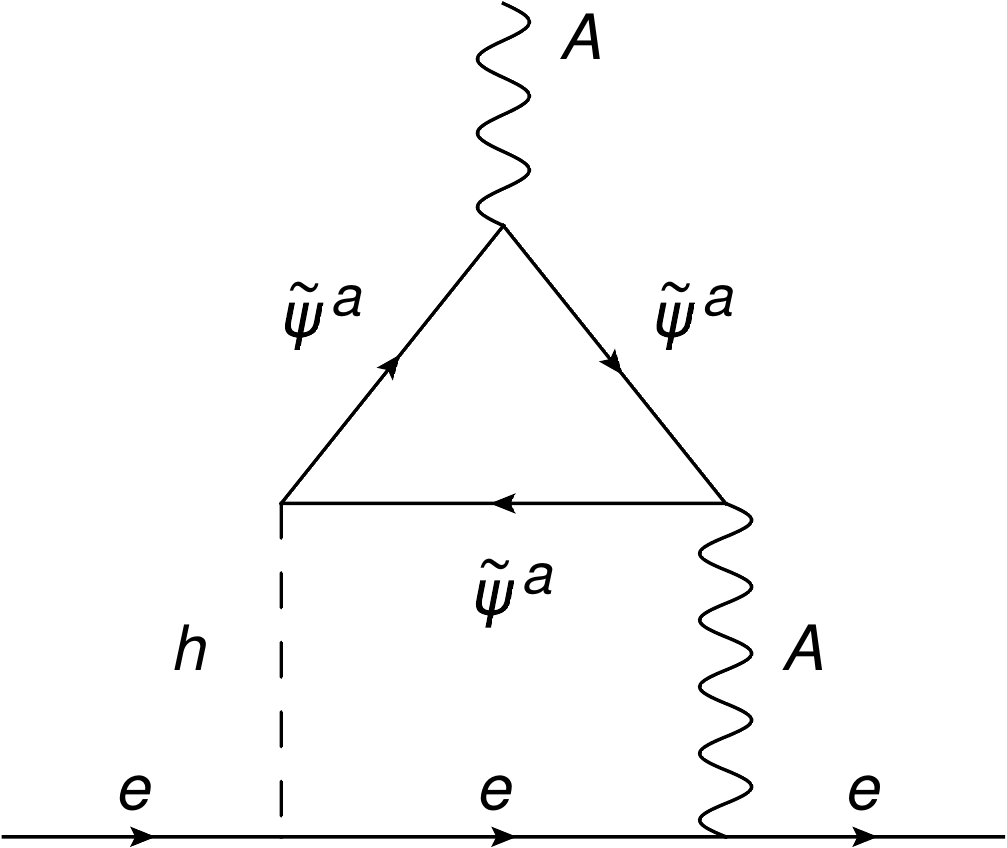}
    \label{fig:dehA}
  \end{subfigure}\;\;
  \captionsetup[subfigure]{justification=centerlast}
  \begin{subfigure}{0.3\textwidth}
    \centering
    \caption{}
    \includegraphics[width=1\textwidth]{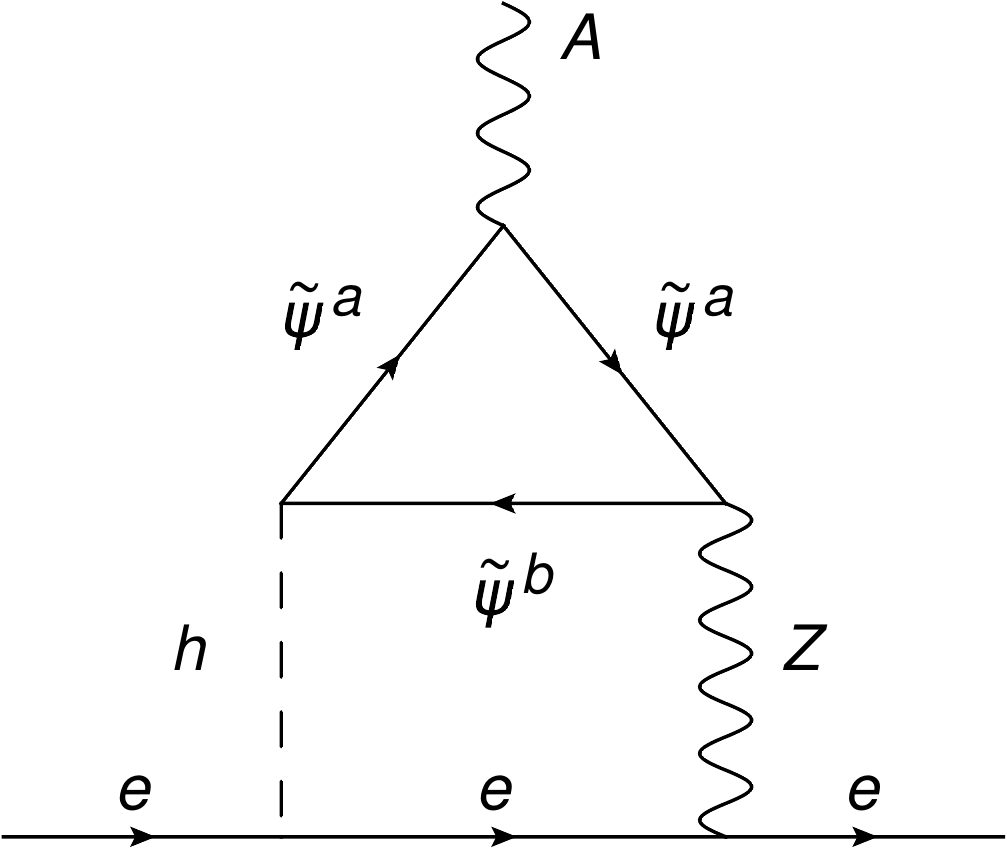}
    \label{fig:dehZ}
  \end{subfigure}\;\;
  \captionsetup[subfigure]{justification=centerlast}
  \begin{subfigure}{0.3\textwidth}
    \centering
    \caption{}
    \includegraphics[width=1\textwidth]{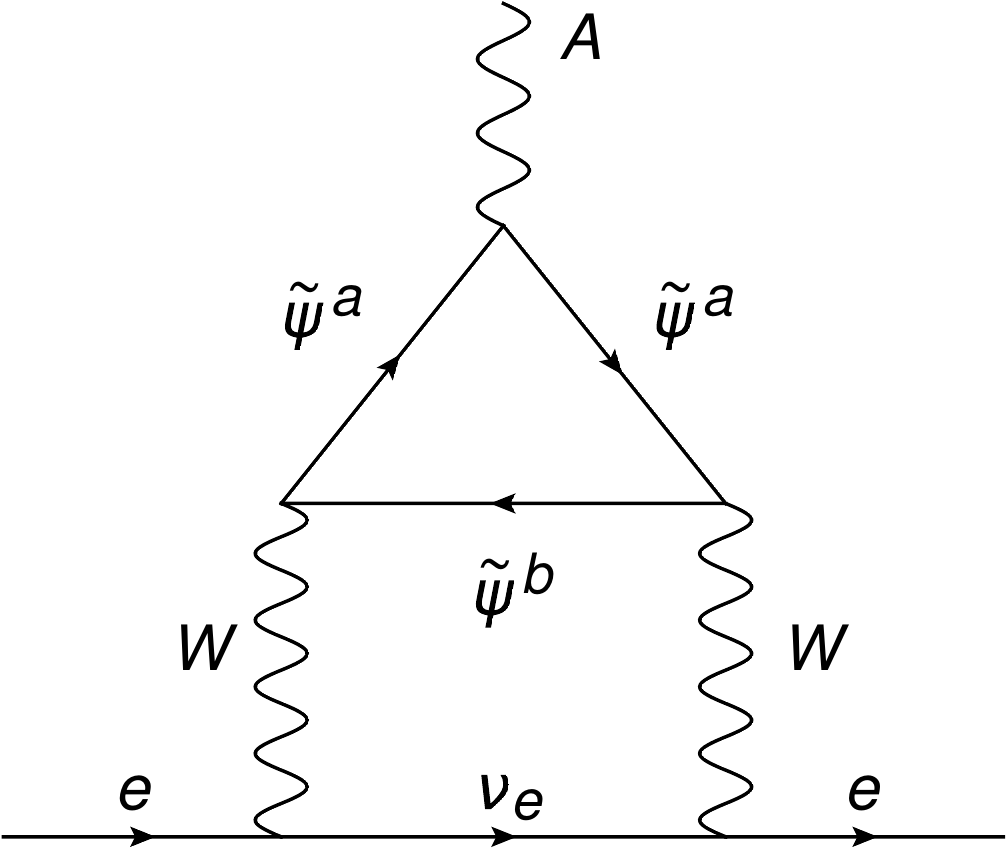}
    \label{fig:deWW}
  \end{subfigure}
\caption{Different examples of diagrams contributing to the EDM of the electron. (a) hA exchange, (b) hZ exchange, and (c) WW exchange.}
\label{fig:de}
%\end{center}
\end{figure}

As explained in Sec.~\ref{sSec:CaseIHiggsdecay}, most constraints from the Higgs signal strengths can be evaded by having almost purely imaginary $\Omega_{aa}$ couplings. This is, however, constrained by the fact that they would contribute to the EDM of the electron through Barr-Zee diagrams, which can easily be computed by adapting results from the literature. First, the Barr-Zee diagram involving the photon and the Higgs boson of Fig.~\ref{fig:dehA} can be computed by adapting the results of Ref.~\cite{Nakai:2016atk}. This diagram and its variations lead to a contribution of
\begin{equation}\label{eq:CaseIEDMPhoton}
  \begin{aligned}
  \frac{d^{Ah}_e}{e} = &-\sum_a \frac{\alpha \tilde{Q}^2_{aa} m_a m_e}{16\pi^3 m_h^2 v}\text{Im}(\Omega_{aa})\\
  &\times\int_0^1 dx \frac{1}{x(1-x)} j\left(0, \frac{m_a^2}{x(1-x)m_h^2}\right),
  \end{aligned}
\end{equation}
where
\begin{equation}\label{eq:CaseIEDMj}
  j(r, s) = \frac{1}{r - s}\left(\frac{r \ln r}{r - 1} - \frac{s \ln s}{s - 1}\right).
\end{equation}

Second, the contribution from the diagrams involving a $Z$ boson and a Higgs boson like that of Fig.~\ref{fig:dehZ} can also be computed using Ref.~\cite{Nakai:2016atk} and lead to
\begin{equation}\label{eq:CaseIEDMZI}
  \begin{aligned}
  \frac{d^{Zh}_e}{e} = \sum_{a, b} \frac{\tilde{Q}_{bb}}{32 \pi^4 m_h^2} g_{ee}^V g_{ee}^S\Bigl(&m_a C_{ab}^1 f_1(m_a, m_b)\\
  + &m_b C_{ab}^2 f_2(m_a, m_b)\Bigr),
  \end{aligned}
\end{equation}
where
\begin{equation}\label{eq:CaseIEDMZII}
  \begin{aligned}
    C_{ab}^1 &=  \text{Re}\left(i g_{ba}^S g_{ab}^A - g_{ba}^P g_{ab}^V\right),\\
    C_{ab}^2 &= -\text{Re}\left(i g_{ba}^S g_{ab}^A + g_{ba}^P g_{ab}^V\right),
  \end{aligned}
\end{equation}
with
\begin{equation}\label{eq:CaseIEDMZIII}
  \begin{aligned}
    g_{ee}^S &= - \frac{m_e}{v}, \quad\quad
    g_{ee}^V = - \frac{\sqrt{g^2 + {g'}^2}}{2}\left(-\frac{1}{2} + 2 s_W^2 \right), \\
    g_{ab}^S &= - \frac{(\Omega_{ba} + \Omega_{ab}^\ast)}{2}, \quad
    g_{ab}^P = -i\frac{(\Omega_{ba} - \Omega_{ab}^\ast)}{2}, \\
    g_{ab}^V &= - \frac{\sqrt{g^2 + {g'}^2}}{2}\left(B_{Rba} + B_{Lba}\right), \\
    g_{ab}^A &= - \frac{\sqrt{g^2 + {g'}^2}}{2}\left(B_{Rba} - B_{Lba}\right),
  \end{aligned}
\end{equation}
and
\begin{equation}\label{eq:CaseIEDMZIV}
  \begin{aligned}
    f_1(m_a, m_b) &= \int_0^1 dx j\left(\frac{m_Z^2}{m_h^2}, \frac{\tilde{\Delta}_{ab}}{m_h^2}\right) , \\
    f_2(m_a, m_b) &= \int_0^1 dx j\left(\frac{m_Z^2}{m_h^2}, \frac{\tilde{\Delta}_{ab}}{m_h^2}\right) \frac{1 - x}{x}, 
  \end{aligned}
\end{equation}
with
\begin{equation}\label{eq:CaseIEDMZV}
  \tilde{\Delta}_{ab} = \frac{x m_a^2 + (1 - x) m_b^2}{x(1 - x)}.
\end{equation}

Third, the contribution from the diagrams involving two $W$ bosons like in Fig.~\ref{fig:deWW} can be computed by adapting the results of Ref.~\cite{Chang:2005ac}. The only subtlety is that the fermions can potentially flow in both directions in the fermion loop. This was not the case in Ref.~\cite{Chang:2005ac}. Thankfully, both diagrams can easily be related and the sum of the two diagrams is
\begin{equation}\label{eq:CaseIEDMWI}
  \begin{aligned}
  &\frac{d_e^{WW}}{e} = -\frac{\alpha^2 m_e}{8\pi^2 s_W^4 m_W^2} \sum_{a, b} \frac{m_a m_b}{m_W^2} \text{Im}\left(\hat{A}_{Lba}\hat{A}_{Rba}^\ast\right)\\
  &\hspace{1.5cm}\times\left[\tilde{Q}_{bb}\mathcal{G}(r_a, r_b, 0) + \tilde{Q}_{aa} \mathcal{G}(r_b, r_a, 0)\right],
  \end{aligned}
\end{equation}
where $r_a = m_a^2/m_W^2$, $r_b = m_b^2/m_W^2$, and
\begin{equation}\label{eq:CaseIEDMWII}
  \begin{aligned}
  &\mathcal{G}(r_a, r_b, r_c) =\\
  &\quad\int_0^1\frac{d\gamma}{\gamma} \int_0^1 dy y\Biggl[\frac{(R - 3 K_{ab})R + 2(K_{ab} + R)y}{4R (K_{ab} - R)^2}\\
  &\hspace{2.7cm}+ \frac{K_{ab}(K_{ab} - 2y)}{2(K_{ab} - R)^3}\ln\frac{K_{ab}}{R}\Biggr],
  \end{aligned}
\end{equation}
with
\begin{equation}\label{eq:CaseIEDMWIII}
  R = y + (1 - y)r_c, \qquad K_{ab} = \frac{r_a}{1 - \gamma} + \frac{r_b}{\gamma}.
\end{equation}
The total EDM is then the sum of Eqs.~\eqref{eq:CaseIEDMPhoton}, \eqref{eq:CaseIEDMZI}, and \eqref{eq:CaseIEDMWI}. In practice, the contribution from Eq.~\eqref{eq:CaseIEDMPhoton} generally overwhelmingly dominates when $\text{BR}(h \to AA')$ is close to its maximally allowed value. In this case, the diagram of Fig.~\ref{fig:dehZ} is suppressed by the fact that $g_{ee}^V$ is small due to an accidental partial cancellation and the diagram of Fig.~\ref{fig:deWW} is suppressed because the Yukawa couplings $A_L$ and $A_R$ are simply much larger than $g$.  The upper limit on the electron EDM that we use is $|d_e| < 4.1 \times 10^{-30} e\;\text{cm}$ at 90\% CL \cite{Roussy:2022cmp}.\footnote{This limit is updated with respect to Ref.~\cite{PhysRevLett.130.141801} which used Ref.~\cite{ACME:2018yjb}, as the result of Ref.~\cite{Roussy:2022cmp} was not available yet. The impact of this new result on our limits is negligible.}

Finally, when $\text{BR}(h \to AA')$ is close to the upper limit that we find, the electron EDM generally constrains $A_L$ and $A_R$ to have a phase difference very close to 0 (or $\pi$). This makes it essentially redundant to consider any other CP-violating observable. 

\subsection{Oblique parameters}\label{sSec:CaseIObliqueParameters}
A sizable $\text{BR}(h \to AA')$ requires some of the Yukawa couplings $A_L$ and $A_R$ to be large. These couplings however have the side effect of causing mixing between fields that are part of different representations of the electroweak gauge groups. This means that a large $\text{BR}(h \to AA')$ is at risk of generating large contributions to the oblique parameters \cite{Peskin:1990zt}. The parameters $S$ and $T$ are computed by using the general results for fermions of Refs.~\cite{Anastasiou:2009rv, Lavoura:1992np, Chen:2003fm, Carena:2007ua, Chen:2017hak, Cheung:2020vqm}. They are given by:
\begin{equation}\label{eq:CaseISTI}
  \begin{aligned}
    S = \frac{1}{2\pi}\sum_{a,b}\Bigl\{&\Bigl(|\hat{A}_{Lab}|^2 +  |\hat{A}_{Rab}|^2\Bigr)\psi_+(y_a, y_b)\\
    &+ 2\text{Re}\Bigl(\hat{A}_{Lab}\hat{A}^\ast_{Rab}\Bigr)\psi_-(y_a, y_b)\\
    & -\frac{1}{2}\Bigl[\Bigl(|X_{ab}|^2 +  |X_{Rab}|^2\Bigr)\chi_+(y_a, y_b)\\
    &+ 2\text{Re}\Bigl(X_{Lab}X^\ast_{Rab}\Bigr)\chi_-(y_a, y_b)\Bigr]\Bigr\},\\
    T = \frac{1}{16\pi s_W^2 c_W^2}&\sum_{a,b}\Bigl\{\Bigl(|\hat{A}_{Lab}|^2 + |\hat{A}_{Rab}|^2\Bigr)\theta_+(y_a, y_b)\\
    &+ 2\text{Re}\Bigr(\hat{A}_{Lab}\hat{A}^\ast_{Rab}\Bigr)\theta_-(y_a, y_b)\\
        & -\frac{1}{2}\Bigl[\Bigl(|X_{Lab}|^2 +  |X_{Rab}|^2\Bigr)\theta_+(y_a, y_b)\\
    &+ 2\text{Re}\Bigl(X_{Lab}X^\ast_{Rab}\Bigr)\theta_-(y_a, y_b)\Bigr]\Bigr\},
  \end{aligned}
\end{equation}
where $y_a = m_a^2/m_Z^2$, $X_{L/R} = -2 B_{L/R} + 2\tilde{Q}s_W^2$ and
\begin{equation}\label{eq:CaseISTII}
  \begin{aligned}
    \psi_+(y_1, y_2)   &= \frac{1}{3} - \frac{1}{9}\ln\frac{y_1}{y_2},\\
    \psi_-(y_1, y_2)   &= -\frac{y_1 + y_2}{6\sqrt{y_1 y_2}},\\
    \chi_+(y_1, y_2)   &= \frac{5(y_1^2 + y_2^2) - 22y_1 y_2}{9(y_1 - y_2)^2} \\
                       &+ \frac{3y_1 y_2 (y_1 + y_2) - y_1^3 - y_2^3}{3(y_1 - y_2)^3}\ln\frac{y_1}{y_2},\\
    \chi_-(y_1, y_2)   &= -\sqrt{y_1 y_2} \Bigl[\frac{y_1 + y_2}{6y_1 y_2} - \frac{y_1 + y_2}{(y_1 - y_2)^2}\\
    &\hspace{1.8cm}+ \frac{2y_1 y_2}{(y_1 - y_2)^3}\ln\frac{y_1}{y_2}\Bigr],\\
    \theta_+(y_1, y_2) &= y_1 + y_2 - \frac{2y_1 y_2}{y_1 - y_2}\ln\frac{y_1}{y_2},\\
    \theta_-(y_1, y_2) &= 2\sqrt{y_1 y_2} \left[\frac{y_1 + y_2}{y_1 - y_2}\ln\frac{y_1}{y_2} - 2\right].
  \end{aligned}
\end{equation}
We use the measurements of the oblique parameters of Ref.~\cite{Zyla:2020zbs} given by
\begin{equation}\label{eq:STconstraints}
  S = 0.00 \pm 0.07, \qquad T = 0.05 \pm 0.06,
\end{equation}
with a correlation of 0.92. We keep points whose $\chi^2$ differ by less than 5.99 from the best fit, which corresponds to 95\% CL limits.

\subsection{Unitarity}\label{sSec:CaseIUnitarity}
As a final constraint, the parameters $A_R$ and $A_L$ are bounded by unitarity. Consider a given scattering between mediators via Higgs exchange and its amplitude $\mathcal{M}$. The latter can be expanded in partial waves as
\begin{equation}\label{eq:CaseIPartialWaves}
  \mathcal{M} = 16\pi\sum_\ell (2\ell + 1) a_\ell P_\ell(\cos\theta),
\end{equation}
where $P_\ell(\cos\theta)$ are the Legendre polynomials.

In the high energy limit, we can work directly with $\psi_1$ and $\psi_2$. It is then simply a question of computing the $a_0$ factor of every possible scattering $\overline{\psi}_1^a\psi_2^b \to \overline{\psi}_1^c\psi_2^d$ for every possible helicity combination. Consider the basis of $\overline{\psi}_1^a\psi_2^b$ pairs given by
\begin{widetext}
\begin{equation}\label{eq:CaseIUnitarityBasisI}
  \overline{\psi}_1^1 \psi_2^1,\;\overline{\psi}_1^1 \psi_2^2,\;...,\;\overline{\psi}_1^1 \psi_2^n,\;
  \overline{\psi}_1^2 \psi_2^1,\;\overline{\psi}_1^2 \psi_2^2,\;...,\;\overline{\psi}_1^2 \psi_2^n,\;...,\;
  \overline{\psi}_1^p \psi_2^1,\;\overline{\psi}_1^p \psi_2^2,\;...,\;\overline{\psi}_1^p \psi_2^n.
\end{equation}
Then the matrix of $a_0$ for the scattering $\overline{\psi}_1^a\psi_2^b \to \overline{\psi}_1^c\psi_2^d$ in the basis of Eq.~\eqref{eq:CaseIUnitarityBasisI} is
\begin{equation}\label{eq:CaseIa0MatrixI}
  a_0^{\text{mat}} = \begin{pmatrix} 
    F_{11}^{11} & F_{11}^{12} & ... & F_{11}^{1n} & F_{11}^{21} & F_{11}^{22} & ... & F_{11}^{2n} & ... & F_{11}^{p1} & F_{11}^{p2} & ... & F_{11}^{pn}\\
    F_{12}^{11} & F_{12}^{12} & ... & F_{12}^{1n} & F_{12}^{21} & F_{12}^{22} & ... & F_{12}^{2n} & ... & F_{12}^{p1} & F_{12}^{p2} & ... & F_{12}^{pn}\\
    ...         & ...         & ... & ...         & ...         & ...         & ... & ...         & ... & ...         & ...         & ... & ...        \\
    F_{1n}^{11} & F_{1n}^{12} & ... & F_{1n}^{1n} & F_{1n}^{21} & F_{1n}^{22} & ... & F_{1n}^{2n} & ... & F_{1n}^{p1} & F_{1n}^{p2} & ... & F_{1n}^{pn}\\
    F_{21}^{11} & F_{21}^{12} & ... & F_{21}^{1n} & F_{21}^{21} & F_{21}^{22} & ... & F_{21}^{2n} & ... & F_{21}^{p1} & F_{21}^{p2} & ... & F_{21}^{pn}\\
    F_{22}^{11} & F_{22}^{12} & ... & F_{22}^{1n} & F_{22}^{21} & F_{22}^{22} & ... & F_{22}^{2n} & ... & F_{22}^{p1} & F_{22}^{p2} & ... & F_{22}^{pn}\\
    ...         & ...         & ... & ...         & ...         & ...         & ... & ...         & ... & ...         & ...         & ... & ...        \\
    F_{2n}^{11} & F_{2n}^{12} & ... & F_{2n}^{1n} & F_{2n}^{21} & F_{2n}^{22} & ... & F_{2n}^{2n} & ... & F_{2n}^{p1} & F_{2n}^{p2} & ... & F_{2n}^{pn}\\
    ...         & ...         & ... & ...         & ...         & ...         & ... & ...         & ... & ...         & ...         & ... & ...        \\
    F_{p1}^{11} & F_{p1}^{12} & ... & F_{p1}^{1n} & F_{p1}^{21} & F_{p1}^{22} & ... & F_{p1}^{2n} & ... & F_{p1}^{p1} & F_{p1}^{p2} & ... & F_{p1}^{pn}\\
    F_{p2}^{11} & F_{p2}^{12} & ... & F_{p2}^{1n} & F_{p2}^{21} & F_{p2}^{22} & ... & F_{p2}^{2n} & ... & F_{p2}^{p1} & F_{p2}^{p2} & ... & F_{p2}^{pn}\\
    ...         & ...         & ... & ...         & ...         & ...         & ... & ...         & ... & ...         & ...         & ... & ...        \\
    F_{pn}^{11} & F_{pn}^{12} & ... & F_{pn}^{1n} & F_{pn}^{21} & F_{pn}^{22} & ... & F_{pn}^{2n} & ... & F_{pn}^{p1} & F_{pn}^{p2} & ... & F_{pn}^{pn}\\
   \end{pmatrix}
\end{equation}
\end{widetext}
where each row corresponds to the same incoming $\overline{\psi}_1^a\psi_2^b$ pair, each column to the same outgoing $\overline{\psi}_1^a\psi_2^b$ pair, and $F_{ab}^{cd}$ is a block given by
\begin{equation}\label{eq:CaseIa0MatrixII}
  F_{ab}^{cd} = \frac{d^{pn}_{ab2} d^{pn}_{cd2}}{32\pi}\begin{pmatrix} -|A_R|^2 & A_R A_L^\ast \\ A_L A_R^\ast & -|A_L|^2 \end{pmatrix}
\end{equation}
and corresponds to $a_0$ for different combinations of helicity in the basis of $(\uparrow \uparrow, \downarrow \downarrow)$. Call $a_0^{\text{eig}}$ the set of eigenvalues of $a_0^{\text{mat}}$. Unitarity then imposes
\begin{equation}\label{eq:Case1a0Limit1}
  \text{max}\left(\left|\text{Re}\left(a_0^{\text{eig}}\right)\right|\right) < \frac{1}{2}.
\end{equation}
This can be verified to reduce to the surprisingly simple requirement that
\begin{equation}\label{eq:Case1a0Limit2}
  |A_R|^2 + |A_L|^2 < \frac{32\pi}{p}.
\end{equation}

\subsection{Results}\label{sSec:CaseIResults}
Having explained how the different constraints are imposed, we now present the limits on $\text{BR}(h \to A A')$ for the fermion case. The parameter space is sampled using a Markov chain with the Metropolis-Hastings algorithm. As the results are only dependent on $Q'$ and $e'$ via the product $Q'e'$, the limits on $\text{BR}(h \to A A')$ are independent of the choice of $Q'$ and we require $|Q'e'| < \sqrt{4\pi}$. To maximize the number of points near the limits and thus reduce the necessary number of simulations, a prior proportional to $\text{BR}(h \to A A')^2$ is assumed. We have verified that the results are independent of the sampling algorithm (and prior), assuming it covers all of the relevant parameter space and sufficient statistics.

\begin{figure}[t!]
\begin{center}
  \includegraphics[width=0.48\textwidth]{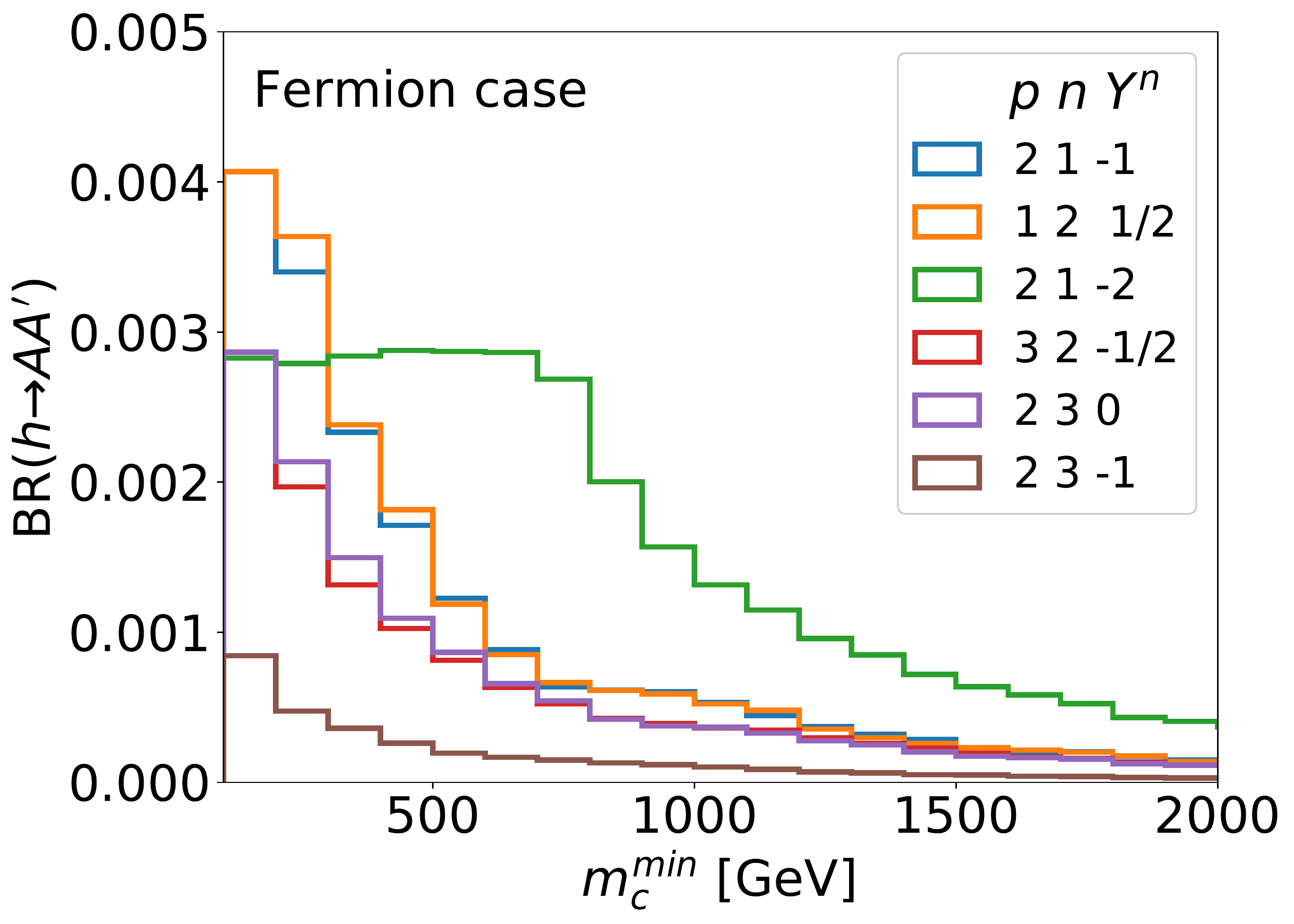}
  \caption{Upper bounds on $\text{BR}(h \to A A')$ for different examples of fermion mediators. The plot does not go below 100~GeV, as LEP bounds prohibit such masses \cite{LEP1, LEP2}. Taken and expanded from Ref.~\cite{PhysRevLett.130.141801}.}
\label{fig:FermionScans}
\end{center}
\end{figure}

The limits on $\text{BR}(h \to A A')$ are shown in Fig.~\ref{fig:FermionScans} for different $p$, $n$, and $Y^n$ as a function of the mass of the lightest electrically charged mediator of mass $m_c^{\text{min}}$. As can be seen, there is an upper bound of about $0.4\%$ that is never crossed. This limit comes from the Higgs signal strengths and is consistent with the estimate of Sec.~\ref{sSec:CaseIHiggsdecay}. For many mediators, the limit on $\text{BR}(h \to A A')$ suddenly starts to drop from its maximal value above a certain threshold in $m_c^{\text{min}}$. This is simply where the constraints from the oblique parameters or unitarity become more powerful than those of the Higgs signal strengths. This is because increasing the masses of the mediators requires larger couplings to maintain a large $\text{BR}(h \to A A')$, which conflicts with these constraints. Some mediators are also sufficiently constrained by other bounds that the lower mass plateau is never reached. Obtaining a large $\text{BR}(h \to A A')$ requires $|Q'e'|$ to be considerably larger than the $|\tilde{Q}_{aa} e|$ of a mediator $\tilde{\psi}^a$. This is difficult for mediators with large electric charges, as $|Q'e'|$ has an upper limit, and is why models with mediators of large electric charges are very constrained. Without the constraints from the electron EDM, the limits on $\text{BR}(h \to A A')$ would be far less stringent as interference terms with the SM contributions could be avoided in the decay of the Higgs to two photons. As explained in Sec.~\ref{sSec:CaseIHiggsdecay}, the electron EDM forces the complex phase to be close to 0 (or $\pi$) which imposes the presence of interference terms.

An important point to mention is that the limits on $\text{BR}(h \to A A')$ depend on the mass of the mediators. In theory, there should be a lower limit on the mass of the charged mediators coming from collider searches. If this mass were higher than the threshold in $m_c^{\text{min}}$, the bound on $\text{BR}(h \to A A')$ would be tightened. In principle, the mediators could decay to exotic channels that have not been probed yet. It is therefore technically impossible to determine a model-independent bound on their masses. However, it would be very difficult for a charged particle of less than a few hundred GeV not to have been observed at the LHC by now. As such, obtaining a large $\text{BR}(h \to A A')$ requires a charged light particle that somehow would have avoided detection.

%%%%%%%%%%%%%%%%%%%%%%%%%%%%%%%%%%%%%%
\section{Scalar mediators}\label{Sec:ScalarMediators}
%%%%%%%%%%%%%%%%%%%%%%%%%%%%%%%%%%%%%%

We now proceed to analyze the four scalar cases of Sec.~\ref{Sec:Assumptions}. With a few exceptions, the treatment is similar to the fermion case up to minor technical details.

\subsection{Field content, Lagrangian, and mass eigenstates}\label{sSec:CaseIILagrangian}
We begin by introducing in more detail the four scalar models. To each model will correspond a series of masses $m_a$, a rotation matrix $R$, and a matrix of Higgs couplings $\Omega$. The results of the subsequent sections will be expressed in terms of these quantities.

\subsubsection*{Scalar case I}
Consider a complex scalar $\phi_1$ that transforms under a representation of $SU(2)_L$ of dimension $p = n \pm 1$, and has a weak hypercharge $Y^p = Y^n + 1/2$ and a charge under $U(1)'$ of $Q'$. Consider another complex scalar $\phi_2$ that transforms under a representation of $SU(2)_L$ of dimension $n$ and has a weak hypercharge of $Y^n$ and a charge under $U(1)'$ of $Q'$. The Lagrangian that controls the masses of the scalars is
\begin{equation}\label{eq:CaseIIaLagrangianI}
  \mathcal{L}_m^1 = -\left[\sum_{a,b,c} \mu \hat{d}_{abc}^{pn} \phi_1^{a\dagger}\phi_2^b H^c + \text{H.c.}\right] - \mu_1^2|\phi_1|^2 - \mu_2^2|\phi_2|^2.
\end{equation}
The $SU(2)_L$ tensor $\hat{d}_{abc}^{pn}$ is given by
\begin{equation}\label{eq:CaseIIadpnabc1}
  \hat{d}^{pn}_{abc} = C^{JM}_{j_1 m_1 j_2 m_2},
\end{equation}
where
\begin{equation}\label{eq:CaseIIadpnabc2}
  \begin{aligned}
    J &= \frac{p -1}{2},       & j_1 &= \frac{n - 1}{2},      & j_2 &= \frac{1}{2}, \\ 
    M &= \frac{p + 1 - 2a}{2}, & m_1 &= \frac{n + 1 - 2b}{2}, & m_2 &=\frac{3 - 2c}{2}.
  \end{aligned}
\end{equation}
The parameter $\mu$ can be made real by a field redefinition. Once the Higgs field obtains a VEV, the Lagrangian $\mathcal{L}_m^1$ will contain the mass terms
\begin{equation}\label{eq:CaseIIaLagrangianII}
  \mathcal{L}_m^1 = -\left[\sum_{a,b}\frac{\mu v}{\sqrt{2}}\hat{d}_{ab2}^{pn} \phi_1^{a\dagger}\phi_2^b + \text{H.c.} \right] - \mu_1^2|\phi_1|^2 - \mu_2^2|\phi_2|^2.
\end{equation}
Let us introduce the notation
\begin{equation}\label{eq:CaseIIaNotationI}
  \hat{\phi} = \begin{pmatrix} \phi_1 \\ \phi_2 \end{pmatrix}
\end{equation}
and
\begin{equation}\label{eq:CaseIIaNotationII}
  \quad d^{pn}_{ab} = \begin{cases} \hat{d}^{pn}_{a(b - p)2}, & \text{if $a \in [1, p]$ and $b \in [p + 1, n + p]$},\\ 0, & \text{otherwise.}\end{cases}\\
\end{equation}
The mass Lagrangian can be written as
\begin{equation}\label{eq:CaseIIaLagrangianIII}
  \mathcal{L}_m^1 \supset -\sum_{a,b} M^2_{ab} \hat{\phi}^{a\dagger} \hat{\phi}^b,
\end{equation}
where the mass matrix is
\begin{equation}\label{eq:CaseIIaMassMatrix}
  M^2 = \begin{pmatrix} \mu^2_1\mathbbm{1}_{p\times p} & 0_{p\times n} \\ 0_{n\times p} & \mu^2_2 \mathbbm{1}_{n\times n} \end{pmatrix}  + \frac{\mu v}{\sqrt{2}} d^{pn}+ \frac{\mu v}{\sqrt{2}} d^{pnT}.
\end{equation}
The mass matrix can be diagonalized by introducing
\begin{equation}\label{eq:CaseIIaFieldRedefinition}
  \hat{\phi} = R \tilde{\phi}.
\end{equation}
The fields $\tilde{\phi}^a$ are the mass eigenstates, and there are $p + n$ of them. Their interactions with the Higgs boson are then described by
\begin{equation}\label{eq:CaseIIaHiggsInteractionsI}
  \mathcal{L}_m^1 \supset -\sum_{a,b} \Omega_{ab} h \tilde{\phi}^{a\dagger} \tilde{\phi}^b,
\end{equation}
where $\Omega$ is a Hermitian matrix given by
\begin{equation}\label{eq:CaseIIaHiggsInteractionsII}
  \Omega = \frac{\mu}{\sqrt{2}}R^\dagger d^{pn}R + \frac{\mu}{\sqrt{2}}R^\dagger d^{pnT}R .
\end{equation}

\subsubsection*{Scalar case II}
Consider a complex scalar $\phi$ that transforms under a representation of $SU(2)_L$ of dimension $n$, and has a weak hypercharge $Y^n$ and a charge under $U(1)'$ of $Q'$. The Lagrangian that controls the masses of the scalars is
\begin{equation}\label{eq:CaseIIbLagrangianI}
  \mathcal{L}_m^2 = -\sum_{r \in \{n - 1, n + 1\}}\sum_{a,b,c,d} \lambda^r \hat{d}_{abcd}^{nr} H^{a\dagger} H^b \phi^{c\dagger}\phi^d - \mu^2|\phi|^2.
\end{equation}
The $SU(2)_L$ tensor $\hat{d}_{abcd}^{nr}$ is given by
\begin{equation}\label{eq:CaseIIbdnabcd1}
  \hat{d}^{nr}_{abcd} = \sum_M C^{JM}_{j_1 m_1 j_2 m_2} C^{JM}_{j_3 m_3 j_4 m_4},
\end{equation}
where $M$ is summed over $\{-J, -J +1, -J + 2, ..., +J\}$ and
\begin{equation}\label{eq:CaseIIbdpnabcd1}
  \begin{aligned}
    j_1 &= \frac{1}{2},      & j_2 &= \frac{n - 1}{2},     & j_3 &= \frac{1}{2},\\
    m_1 &= \frac{3 - 2a}{2}, & m_2 &= \frac{n + 1 - 2c}{2}, & m_3 &= \frac{3 - 2b}{2},\\ 
    j_4 &= \frac{n - 1}{2},      & J &= \frac{r - 1}{2},\\
    m_4 &= \frac{n + 1 - 2d}{2}.
  \end{aligned}
\end{equation}
There are generally two possible ways to contract the $SU(2)_L$ indices, and each possible contraction is taken into account by its own coefficient $\lambda^r$. The only exception to this is when $\phi$ is a singlet, in which case only the $r = n + 1$ term leads to a non-zero $\hat{d}^{nr}_{abcd}$ tensor. The parameters $\lambda^r$ are always real. Once the Higgs obtains a VEV, the Lagrangian $\mathcal{L}_m^2$ will contain the mass terms
\begin{equation}\label{eq:CaseIIbLagrangianII}
  \mathcal{L}_m^2 \supset -\sum_{r \in \{n - 1, n + 1\}}\sum_{c,d} \frac{\lambda^r v^2}{2} \hat{d}_{22cd}^{nr}\phi^{c\dagger}\phi^d - \mu^2|\phi|^2.
\end{equation}
With the notation
\begin{equation}\label{eq:CaseIIbNotationI}
  \quad d^{nr}_{ab} = \hat{d}^{nr}_{22ab},
\end{equation}
the mass Lagrangian can be written as
\begin{equation}\label{eq:CaseIIbLagrangianIII}
  \mathcal{L}_m^2 \supset -\sum_{a,b} M^2_{ab} \phi^{a\dagger} \phi^b,
\end{equation}
where the mass matrix is
\begin{equation}\label{eq:CaseIIbMassMatrix}
  M^2 = \mu^2 + \sum_{r \in \{n - 1, n + 1\}}\frac{\lambda^r v^2}{2}d^{nr}.
\end{equation}
The mass matrix can be diagonalized by introducing
\begin{equation}\label{eq:CaseIIbFieldRedefinition}
  \phi = R \tilde{\phi}.
\end{equation}
The fields $\tilde{\phi}^a$ are the mass eigenstates, and there are $n$ of them. Their interactions with the Higgs boson are then described by
\begin{equation}\label{eq:CaseIIbHiggsInteractionsI}
  \mathcal{L}_m^2 \supset -\sum_{a,b} \Omega_{ab} h \tilde{\phi}^{a\dagger} \tilde{\phi}^b,
\end{equation}
where $\Omega$ is a real diagonal matrix given by
\begin{equation}\label{eq:CaseIIbHiggsInteractionsII}
  \Omega = \sum_{r \in \{n - 1, n + 1\}} \lambda^r v R^\dagger d^{nr} R.
\end{equation}
We note that the mixing matrix could simply be taken as the identity in this case, as there are never multiple states with the same electric charge. We keep $R$ to maintain a uniform notation and order the particles by mass.

\subsubsection*{Scalar case III}
Consider a complex scalar $\phi_1$ that transforms under a representation of $SU(2)_L$ of dimension $p \in \{n - 2, n, n + 2\}$, and has a weak hypercharge $Y^p = Y^n$ and a charge under $U(1)'$ of $Q'$. Consider another complex scalar $\phi_2$ that transforms under a representation of $SU(2)_L$ of dimension $n$ and has a weak hypercharge $Y^n$ and a charge under $U(1)'$ of $Q'$. The Lagrangian that controls the masses of the scalars is
\begin{equation}\label{eq:CaseIIcLagrangianI}
  \begin{aligned}
  \mathcal{L}_m^3 = &-\left[\sum_{r \in \mathcal{R}}\sum_{a,b,c,d} \lambda^r \hat{d}_{abcd}^{pnr} H^{a\dagger} H^b \phi^{c\dagger}_1\phi^d_2 + \text{H.c.}\right]\\
  &- \mu_1^2|\phi_1|^2 - \mu_2^2|\phi_2|^2,
  \end{aligned}
\end{equation}
where $\mathcal{R}=\{n - 1, n + 1\}\cap \{p - 1, p + 1\}$. The $SU(2)_L$ tensor $\hat{d}_{abcd}^{pnr}$ is given by
\begin{equation}\label{eq:CaseIIcdnabcd1}
  \hat{d}^{pnr}_{abcd} = \sum_M C^{JM}_{j_1 m_1 j_2 m_2} C^{JM}_{j_3 m_3 j_4 m_4},
\end{equation}
where $M$ is summed over $\{-J, -J + 1, -J + 2, ..., +J\}$ and
\begin{equation}\label{eq:CaseIIcdpnabcd1}
  \begin{aligned}
    j_1 &= \frac{1}{2},      & j_2 &= \frac{p - 1}{2},     & j_3 &= \frac{1}{2},\\
    m_1 &= \frac{3 - 2a}{2}, & m_2 &= \frac{p + 1 - 2c}{2} & m_3 &= \frac{3 - 2b}{2},\\
    j_4 &= \frac{n - 1}{2},      & J &= \frac{r - 1}{2},\\
    m_4 &= \frac{n + 1 - 2d}{2}.
  \end{aligned}
\end{equation}
If $p = n$, there are two possible contractions of the $SU(2)_L$ indices, the only exception being if $\phi_1$ and $\phi_2$ are both singlets in which case only the $r = p + 1$ term leads to a non-zero $\hat{d}^{pnr}_{abcd}$. When $p$ and $q$ differ by 2, only one term is allowed. One $\lambda^r$ can be made real by field redefinition. Therefore, there will be a complex phase that cannot generally be reabsorbed when $p = n$, but not otherwise. Once the Higgs field obtains a VEV, the Lagrangian $\mathcal{L}_m^3$ will contain the mass terms
\begin{equation}\label{eq:CaseIIcLagrangianII}
  \begin{aligned}
  \mathcal{L}_m^3 \supset &-\left[\sum_{r \in \mathcal{R}}\sum_{c,d} \frac{\lambda^r v^2}{2} \hat{d}_{22cd}^{pnr}\phi^{c\dagger}_1\phi^d_2 + \text{H.c.}\right]\\
  &- \mu_1^2|\phi_1|^2 - \mu_2^2|\phi_2|^2.
  \end{aligned}
\end{equation}
Introduce the notation
\begin{equation}\label{eq:CaseIIcNotationI}
  \hat{\phi} = \begin{pmatrix} \phi_1 \\ \phi_2 \end{pmatrix}
\end{equation}
and
\begin{equation}\label{eq:CaseIIcNotationII}
  \quad d^{pnr}_{ab} = \begin{cases} \hat{d}^{pnr}_{22a(b - p)}, & \text{if $a \in [1, p]$ and $b \in [p + 1, n + p]$},\\ 0, & \text{otherwise.}\end{cases}\\
\end{equation}
The mass Lagrangian can be written as
\begin{equation}\label{eq:CaseIIcLagrangianIII}
  \mathcal{L}_m^3 \supset -\sum_{a,b} M^2_{ab} \hat{\phi}^{a\dagger} \hat{\phi}^b,
\end{equation}
where the mass matrix is
\begin{equation}\label{eq:CaseIIcMassMatrix}
\begin{aligned}
  &M^2 =\\
  &\;\begin{pmatrix} \mu^2_1\mathbbm{1}_{p\times p} & 0_{p\times n} \\ 0_{n\times p} & \mu^2_2\mathbbm{1}_{n\times n} \end{pmatrix} + 
  \sum_{r \in \mathcal{R}}\left[\frac{\lambda^r v^2}{2} d^{pnr}+ \frac{\lambda^{r\ast} v^2}{2} d^{pnrT}\right].
\end{aligned}
\end{equation}
The mass matrix can be diagonalized by introducing
\begin{equation}\label{eq:CaseIIcFieldRedefinition}
  \hat{\phi} = R \tilde{\phi}.
\end{equation}
The fields $\tilde{\phi}^a$ are the mass eigenstates, and there are $p + n$ of them. Their interactions with the Higgs boson are then described by
\begin{equation}\label{eq:CaseIIcHiggsInteractionsI}
  \mathcal{L}_m^3 \supset -\sum_{a,b} \Omega_{ab} h \tilde{\phi}^{a\dagger} \tilde{\phi}^b,
\end{equation}
where $\Omega$ is a Hermitian matrix given by
\begin{equation}\label{eq:CaseIIcHiggsInteractionsII}
  \Omega = \sum_{r \in \mathcal{R}}\left[\lambda^r vR^\dagger d^{pnr}R+ \lambda^{r\ast} v R^\dagger d^{pnrT}R\right].
\end{equation}

\subsubsection*{Scalar case IV}
Consider a complex scalar $\phi_1$ that transforms under a representation of $SU(2)_L$ of dimension $p \in \{n - 2, n, n + 2\}$, and has a weak hypercharge $Y^p = Y^n + 1$ and a charge under $U(1)'$ of $Q'$. Consider another complex scalar $\phi_2$ that transforms under a representation of $SU(2)_L$ of dimension $n$ and has a weak hypercharge $Y^n$ and a charge under $U(1)'$ of $Q'$. Assume that $n$ and $p$ are not both 1. The Lagrangian that controls the masses of the scalars is
\begin{equation}\label{eq:CaseIIdLagrangianI}
  \mathcal{L}_m^4 = -\left[\lambda \hat{d}_{abcd}^{pn} H^a H^b \phi^{c\dagger}_1\phi^d_2 + \text{H.c.}\right] - \mu_1^2|\phi_1|^2 - \mu_2^2|\phi_2|^2.
\end{equation}
The $SU(2)_L$ tensor $\hat{d}_{abcd}^{pn}$ is given by
\begin{equation}\label{eq:CaseIIddnabcd1}
  \hat{d}^{pn}_{abcd} = \sum_{M_1} C^{J_1 M_1}_{j_1 m_1 j_2 m_2} C^{J_2 M_2}_{J_1 M_1 j_3 m_3},
\end{equation}
where $M_1$ is summed over $\{-1, 0, 1\}$ and
\begin{equation}\label{eq:CaseIIddpnabcd1}
  \begin{aligned}
    j_1 &= \frac{1}{2},      & j_2 &= \frac{1}{2},     & j_3 &= \frac{n - 1}{2},\\
    m_1 &= \frac{3 - 2a}{2}, & m_2 &= \frac{3 - 2b}{2} & m_3 &= \frac{n + 1 - 2d}{2},\\
    J_1 &= 1, & J_2 &= \frac{p - 1}{2},\\
        &     & M_2 &= \frac{p + 1 -2c}{2}.
  \end{aligned}
\end{equation}
There is only a single possible contraction, as the two Higgs doublets can only be combined in a single non-trivial way. This does not occur for cases~II and III because they contain both the Higgs doublet and its conjugate. Because there is only one coefficient, $\lambda$ can always be made real by a field redefinition. Once the Higgs field obtains a VEV, the Lagrangian $\mathcal{L}_m^4$ will contain the mass terms
\begin{equation}\label{eq:CaseIIdLagrangianII}
  \mathcal{L}_m^4 \supset -\left[\sum_{c,d} \frac{\lambda v^2}{2} \hat{d}_{22cd}^{pn}\phi^{c\dagger}_1\phi^d_2 + \text{H.c.}\right] - \mu_1^2|\phi_1|^2 - \mu_2^2|\phi_2|^2.
\end{equation}
Introduce the notation
\begin{equation}\label{eq:CaseIIdNotationI}
  \hat{\phi} = \begin{pmatrix} \phi_1 \\ \phi_2 \end{pmatrix}
\end{equation}
and
\begin{equation}\label{eq:CaseIIdNotationII}
  \quad d^{pn}_{ab} = \begin{cases} \hat{d}^{pn}_{22a(b - p)}, & \text{if $a \in [1, p]$ and $b \in [p + 1, n + p]$},\\ 0, & \text{otherwise.}\end{cases}\\
\end{equation}
The mass Lagrangian can be written as
\begin{equation}\label{eq:CaseIIdLagrangianIII}
  \mathcal{L}_m^4 \supset -\sum_{a,b} M^2_{ab} \hat{\phi}^{a\dagger} \hat{\phi}^b,
\end{equation}
where the mass matrix is
\begin{equation}\label{eq:CaseIIdMassMatrix}
  M^2 = \begin{pmatrix} \mu^2_1\mathbbm{1}_{p\times p} & 0_{p\times n} \\ 0_{n\times p} & \mu^2_2 \mathbbm{1}_{n\times n} \end{pmatrix} +
  \frac{\lambda v^2}{2}\left[d^{pn} +  d^{pnT}\right].
\end{equation}
The mass matrix can be diagonalized by introducing
\begin{equation}\label{eq:CaseIIdFieldRedefinition}
  \hat{\phi} = R \tilde{\phi}.
\end{equation}
The fields $\tilde{\phi}^a$ are the mass eigenstates, and there are $p + n$ of them. Their interactions with the Higgs boson are then described by
\begin{equation}\label{eq:CaseIIdHiggsInteractionsI}
  \mathcal{L}_m^4 \supset -\sum_{a,b} \Omega_{ab} h \tilde{\phi}^{a\dagger} \tilde{\phi}^b,
\end{equation}
where $\Omega$ is a Hermitian matrix given by
\begin{equation}\label{eq:CaseIIdHiggsInteractionsII}
  \Omega = \lambda v R^\dagger d^{pn}R + \lambda v R^\dagger d^{pnT}R.
\end{equation}

\subsection{Gauge interactions}\label{sSec:CaseIIInteractions}
The interactions of $A/A'$ with $\tilde{\phi}^a$ are controlled by
\begin{equation}\label{eq:CaseIIPhotonInteractionsI}
  \begin{aligned}
  \mathcal{L}_g \supset & \Bigl(ie A_\mu \partial^\mu \tilde{\phi}^\dagger \tilde{Q} \tilde{\phi} + i Q'e'A'_\mu \partial^\mu \tilde{\phi}^\dagger \tilde{\phi} + \text{H.c.}\Bigr)\\
                & +       \Bigl(e^2 A_\mu  A^\mu    \tilde{\phi}^\dagger \tilde{Q}^2 \tilde{\phi} + Q'{}^2e'{}^2 A'_\mu A'{}^\mu \tilde{\phi}^\dagger \tilde{\phi} \\
                & \hspace{0.5cm}+ 2 Q' e' e A'_\mu A^\mu\tilde{\phi}^\dagger \tilde{Q} \tilde{\phi} \Bigr),
  \end{aligned}
\end{equation}
where $\tilde{Q}$ is a diagonal charge matrix given by
\begin{equation}\label{eq:CaseIIPhotonInteractionsII}
  \tilde{Q} = R^\dagger \hat{Q} R, 
\end{equation}
and
\begin{equation}\label{eq:CaseIIPhotonInteractionsIII}
\begin{aligned}
  &\text{Cases I, III, IV:}&\ &\hat{Q} = \begin{pmatrix} Y^p + T_3^p & 0_{p\times n} \\ 0_{n\times p} & Y^n + T_3^n \end{pmatrix},\\
  &\text{Case II:}& &\hat{Q} = Y^n + T_3^n, 
\end{aligned}
\end{equation}
with $(T_3^p)_{ab} = (p + 1 - 2a)\delta_{ab}/2$ and similarly for $T_3^n$.

The interactions between the $Z$ boson and $\tilde{\phi}^a$ are controlled by the terms
\begin{equation}\label{eq:CaseIIZInteractionsI}
  \begin{aligned}
  \mathcal{L}_g \supset &\left(i \sqrt{g^2 + {g'}^2} Z_\mu \partial^\mu \tilde{\phi}^\dagger B \tilde{\phi} + \text{H.c.}\right)\\
  & + \left(g^2 + {g'}^2\right) Z_\mu Z^\mu \tilde{\phi}^\dagger B^2 \tilde{\phi},
  \end{aligned}
\end{equation}
where $B$ is Hermitian but, in general, non-diagonal and given by
\begin{equation}\label{eq:CaseIIZInteractionsII}
  \begin{aligned}
    &\text{Cases I, III, IV:}\\
    & B = R^\dagger \begin{pmatrix} -s_W^2 Y^p + c_W^2 T_3^p & 0_{p\times n} \\ 0_{n\times p} & -s_W^2 Y^n + c_W^2 T_3^n \end{pmatrix}R,\\
    &\text{Case II:}\\
    & B = R^\dagger \left(-s_W^2 Y^n + c_W^2 T_3^n\right)R.
  \end{aligned}
\end{equation}

The interaction among $\tilde{\phi}^a$, a $A/A'$ boson, and a $Z$ boson is controlled by the terms
\begin{equation}\label{eq:CaseIIZAInteraction}
  \begin{aligned}
  \mathcal{L}_g \supset &2 e \sqrt{g^2 + {g'}^2} A_\mu Z^\mu \tilde{\phi}^\dagger \tilde{Q}B \tilde{\phi} \\
  & + 2 Q' e' \sqrt{g^2 + {g'}^2} A'_\mu Z^\mu \tilde{\phi}^\dagger B \tilde{\phi},
  \end{aligned}
\end{equation}
where we note that $[\tilde{Q},B]=0$.

The interactions between the $W$ boson and $\tilde{\phi}^a$ is controlled by the terms
\begin{equation}\label{eq:CaseIIWInteractionsI}
  \begin{aligned}
    \mathcal{L}_g \supset & \frac{i g}{\sqrt{2}}W^+_\mu\Bigl(\partial^\mu \tilde{\phi}^\dagger \hat{A}\tilde{\phi} - \tilde{\phi}^\dagger \hat{A}\partial^\mu \tilde{\phi}\Bigr) + \text{H.c.}\\
                  & +\frac{g^2}{2}\Bigl(W^+_\mu W^{+\mu}\tilde{\phi}^\dagger \hat{A}^2\tilde{\phi} + W^+_{\mu} W^{-\mu}\tilde{\phi}^\dagger \hat{A}\hat{A}^\dagger \tilde{\phi}\\
                  & \hspace{0.7cm}+ W^+_{\mu} W^{-\mu}\tilde{\phi}^\dagger \hat{A}^\dagger\hat{A} \tilde{\phi} + W^-_\mu W^{-\mu}\tilde{\phi}^\dagger \hat{A}^\dagger{}^2\tilde{\phi}\Bigr),
  \end{aligned}
\end{equation}
where
\begin{equation}\label{eq:CaseIIWInteractionsII}
  \begin{aligned}
  &\text{Cases I, III, IV:}
  &&\hat{A} = R^\dagger \begin{pmatrix}T_+^p & 0_{p\times n} \\ 0_{n\times p} & T_+^n \end{pmatrix} R,\\
  &\text{Case II:}
  &&\hat{A} = R^\dagger T_+^n R,
  \end{aligned}
\end{equation}
with $(T_+^p)_{ab} = \sqrt{a(p - a)}\delta_{a,b-1}$ and similarly for $T_+^n$.

\subsection{Relevant Higgs decays}\label{sSec:CaseIIHiggsdecay}
\begin{figure}[t!]
\begin{center}
 \captionsetup[subfigure]{justification=centerlast}
 \begin{subfigure}{0.4\textwidth}
    \centering
    \caption{}
    \includegraphics[width=1\textwidth]{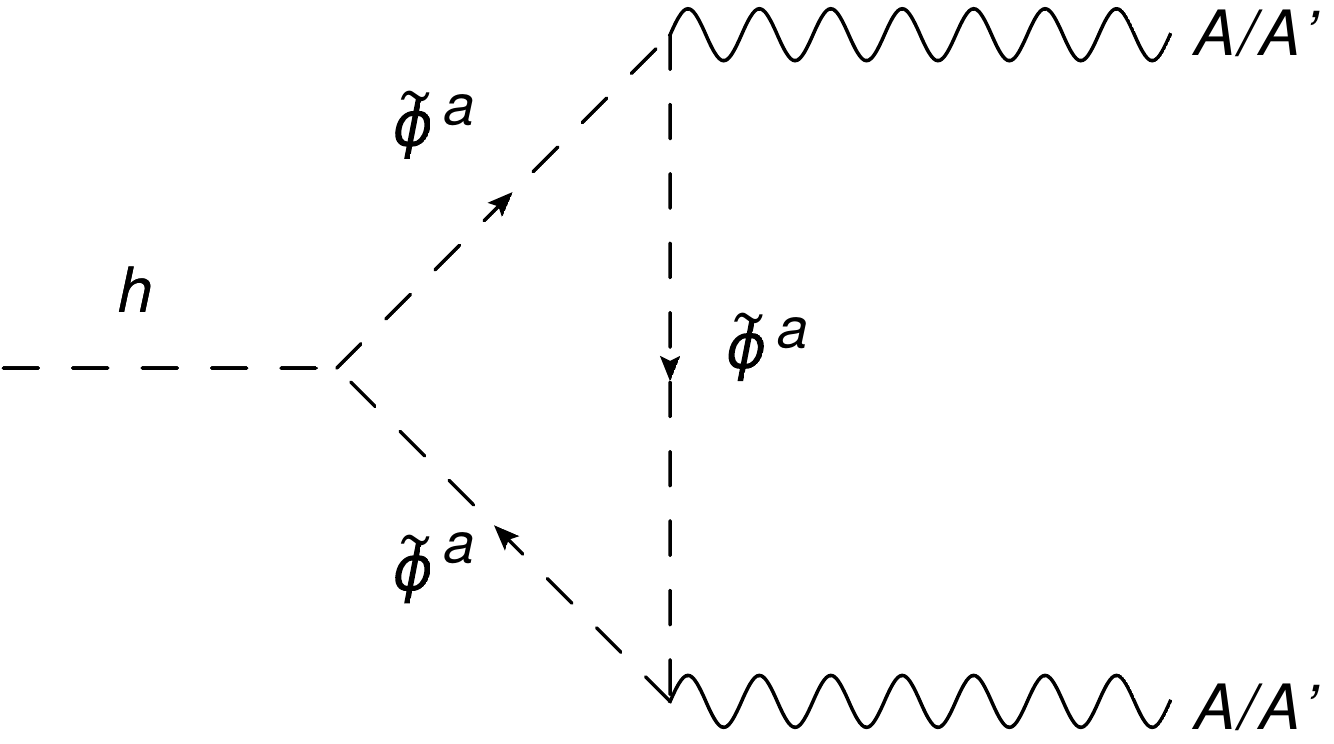}
    \label{fig:CaseIIHiggsDecay3}
  \end{subfigure}\qquad
\captionsetup[subfigure]{justification=centerlast}
  \begin{subfigure}{0.4\textwidth}
    \centering
    \caption{}
    \includegraphics[width=1\textwidth]{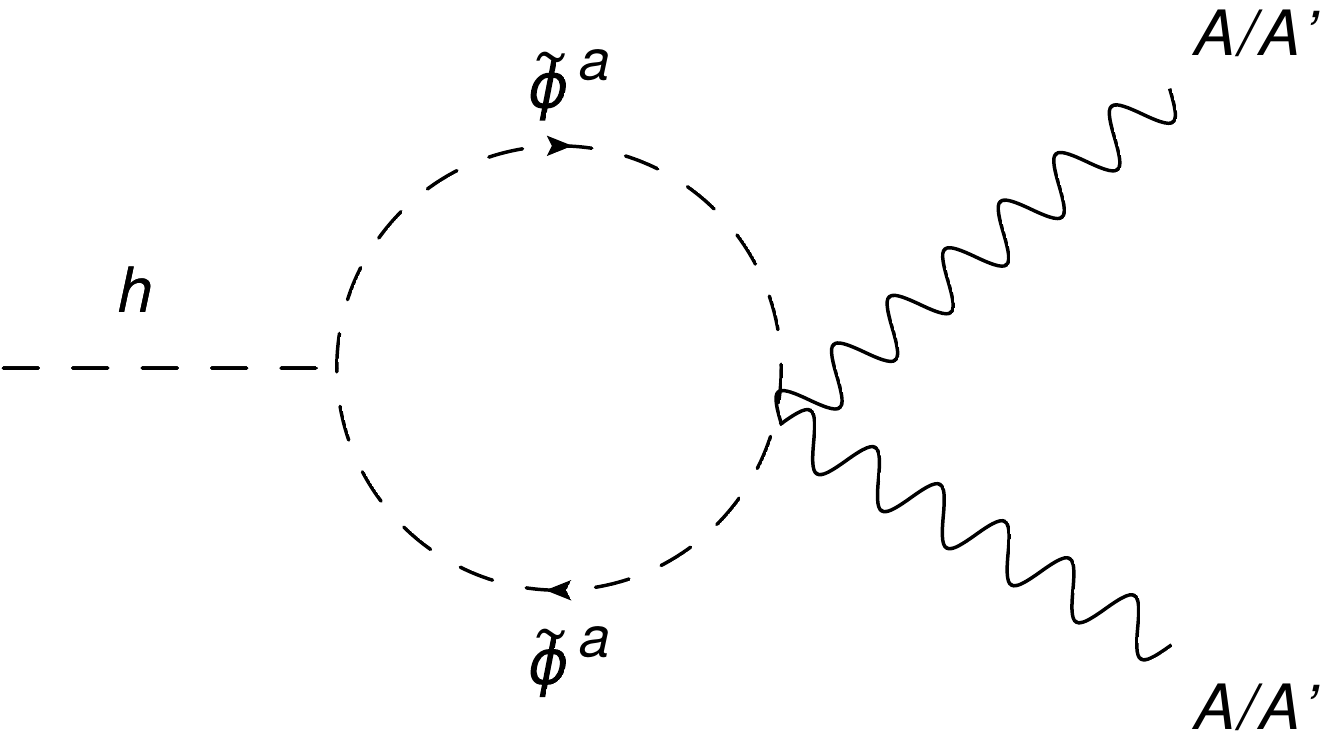}
    \label{fig:CaseIIHiggsDecay5}
  \end{subfigure}
 \captionsetup[subfigure]{justification=centerlast}
 \begin{subfigure}{0.4\textwidth}
    \centering
    \caption{}
    \includegraphics[width=1\textwidth]{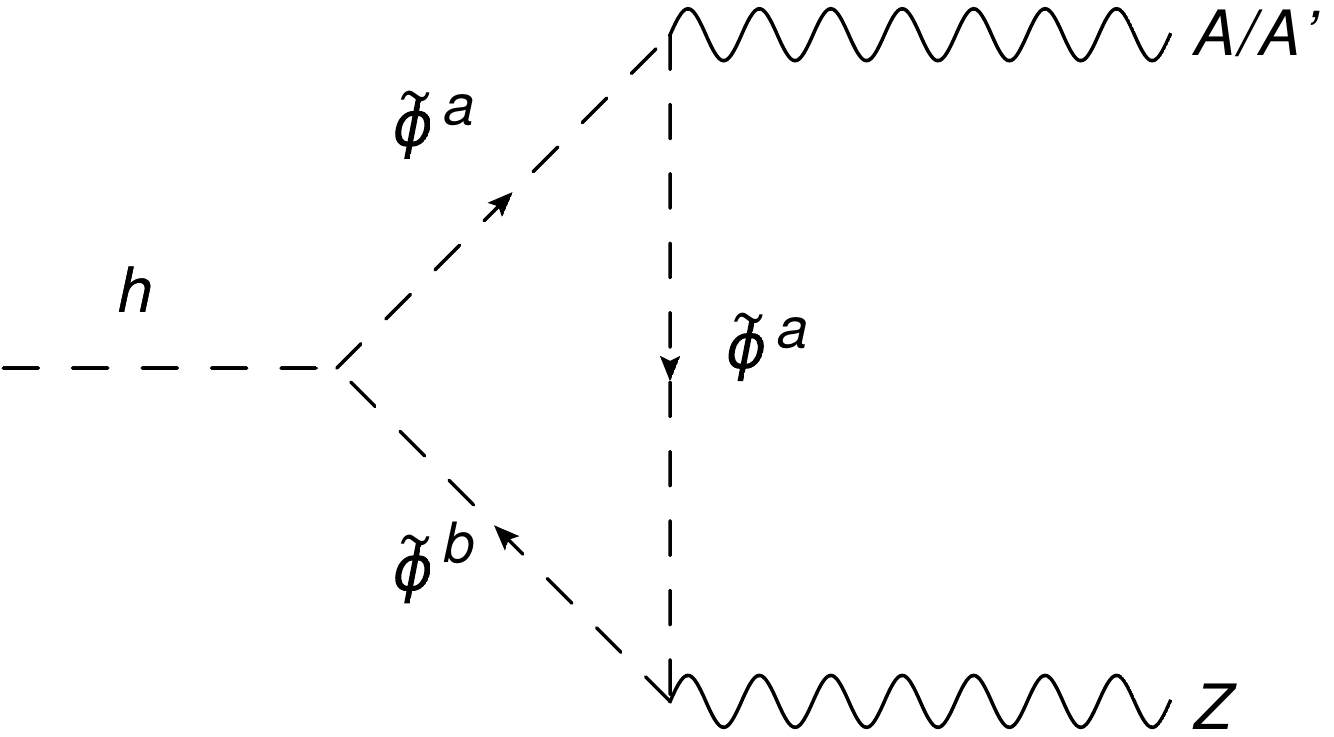}
    \label{fig:CaseIIHiggsDecay4}
  \end{subfigure}\qquad
  \captionsetup[subfigure]{justification=centerlast}
  \begin{subfigure}{0.4\textwidth}
    \centering
    \caption{}
    \includegraphics[width=1\textwidth]{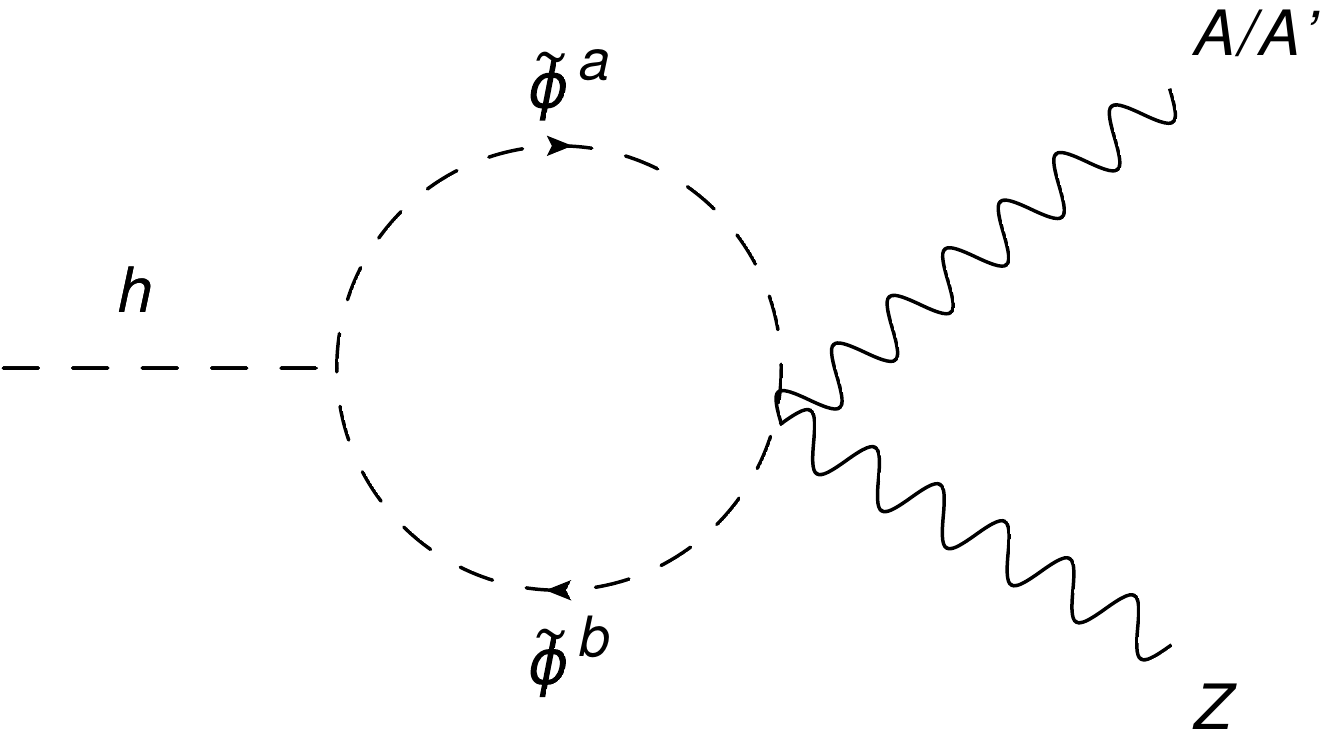}
    \label{fig:CaseIIHiggsDecay6}
  \end{subfigure}
\caption{(a) and (b) Higgs decay to two $A/A'$. (c) and (d) Higgs decay to a $Z$ boson and $A/A'$. Diagrams with the mediator flow inverted also exist for (a) and (c).
}
\label{fig:CaseIIHiggsDecay}
\end{center}
\end{figure}

The mediators $\tilde{\phi}^a$ lead to an amplitude for the Higgs decaying to $AA$, $AA'$ and $A'A'$. The relevant diagrams are shown in Figs.~\ref{fig:CaseIIHiggsDecay3} and \ref{fig:CaseIIHiggsDecay5}. Labelling the momenta of the two gauge bosons as $p_1$ and $p_2$, the amplitudes once again take the form
\begin{equation}\label{eq:CaseIIHiggsAAdecayI}
  \begin{aligned}
    M^{h\to AA} = & S^{h\to AA} \left(p_1\cdot p_2 g_{\mu\nu} - p_{1\mu} p_{2\nu}\right)\epsilon^\nu_{p_1}\epsilon^\mu_{p_2}\\
                & + i\tilde{S}^{h\to AA}\epsilon_{\mu\nu\alpha\beta}p_1^\alpha p_2^\beta\epsilon^\nu_{p_1}\epsilon^\mu_{p_2},\\
    M^{h\to AA'} = & S^{h\to AA'} \left(p_1\cdot p_2 g_{\mu\nu} - p_{1\mu} p_{2\nu}\right)\epsilon^\nu_{p_1}\epsilon^\mu_{p_2}\\
                & + i\tilde{S}^{h\to AA'}\epsilon_{\mu\nu\alpha\beta}p_1^\alpha p_2^\beta\epsilon^\nu_{p_1}\epsilon^\mu_{p_2},\\
    M^{h\to A'A'} = & S^{h\to A'A'} \left(p_1\cdot p_2 g_{\mu\nu} - p_{1\mu} p_{2\nu}\right)\epsilon^\nu_{p_1}\epsilon^\mu_{p_2}\\
                & + i\tilde{S}^{h\to A'A'}\epsilon_{\mu\nu\alpha\beta}p_1^\alpha p_2^\beta\epsilon^\nu_{p_1}\epsilon^\mu_{p_2},
  \end{aligned}
\end{equation}
with the coefficients now given at one loop by
\begin{equation}\label{eq:CaseIIHiggsAAdecayII}
  \begin{aligned}
    S^{h\to AA  } &= e^2    \sum_a \Omega_{aa} \tilde{Q}_{aa}^2  S_a + S^{h\to AA}_\text{SM},\\
    S^{h\to AA' } &= e e'   \sum_a \Omega_{aa} \tilde{Q}_{aa} Q' S_a,\\
    S^{h\to A'A'} &= {e'}^2 \sum_a \Omega_{aa} {Q'}^2            S_a, \\
    \tilde{S}^{h\to AA  } &= \tilde{S}^{h\to AA}_\text{SM}, \quad \tilde{S}^{h\to AA' } = 0, \quad
    \tilde{S}^{h\to A'A'} = 0,
  \end{aligned}
\end{equation}
where $S^{h\to AA}_\text{SM}$ and $\tilde{S}^{h\to AA}_\text{SM}$ are the SM contributions to their respective coefficients and
\begin{equation}\label{eq:CaseIIHiggsAAdecayIII}
  S_a = \frac{1}{4\pi^2 m_h^2}\left[ 1 + 2m_a^2 C_0(0, 0, m_h^2; m_a, m_a, m_a) \right].
\end{equation}
The decay of the Higgs boson to a $Z$ boson and either $A$ or $A'$ is shown in Figs.~\ref{fig:CaseIIHiggsDecay4} and \ref{fig:CaseIIHiggsDecay6} and has a similar form to Eq.~\eqref{eq:CaseIIHiggsAAdecayI}, albeit with far more complicated coefficients.

We reiterate that none of these models contribute to $\tilde{S}^{h\to AA}$. In addition, the contributions to $S^{h\to AA  }$ are all forced to be real. It therefore means that these models will unavoidably lead to interference terms with the SM contributions. As such, they will not be able to circumvent the constraints of the Higgs signal strengths like the fermion mediators potentially could have. As such, we will not study the electron EDM for the scalar models.

\subsection{Higgs signal strengths}\label{sSec:CaseIIHiggsSS}
The constraints associated with the Higgs signal strengths are applied in the same way as for the fermion case.

\subsection{Oblique parameters}\label{sSec:CaseIIObliqueParameters}
For the scalar models, we have performed the computation of the oblique parameters and obtained:
\begin{equation}\label{eq:CaseIISTI}
  \begin{aligned}
    S &= \frac{1}{2\pi}\sum_{a,b}\Bigl[|B_{ab}|^2 - (c_W^2 - s_W^2)B_{ab}\tilde{Q}_{ab} - c_W^2 s_W^2 \tilde{Q}_{ab}^2 \Bigr]\\
    &\hspace{1.5cm}\times F_1(y_a, y_b),\\
    T &= \frac{1}{16\pi c_W^2 s_W^2}\\
    &\hspace{0.2cm}\times\Biggl[\sum_{a,b}|\hat{A}_{ab}|^2 F_2(y_a, y_b) - \sum_a \left[\hat{A}\hat{A}^\dagger + \hat{A}^\dagger \hat{A}\right]_{aa} F_3(y_a)\\
      & \hspace{0.7cm} -2 \sum_{a,b}|B_{ab}|^2 F_2(y_a, y_b) + 4 \sum_a \left[B^2\right]_{aa} F_3(y_a) \Biggr],
  \end{aligned}
\end{equation}
where $y_a = m_a^2/m_Z^2$ and
\begin{equation}\label{eq:CaseIISTII}
  \begin{aligned}
    F_1(y_1, y_2) &= -\frac{5y_1^2 - 22 y_1 y_2 + 5y_2^2}{9(y_1 - y_2)^2}\\
    &\hspace{0.4cm}+ \frac{2\left(y_1^2(y_1 - 3y_2)\ln y_1 - y_2^2(y_2 - 3y_1)\ln y_2 \right)}{3(y_1 - y_2)^3},\\
    F_2(y_1, y_2) &= 3\left(y_1 + y_2\right) - \frac{2\left(y_1^2\ln y_1 - y_2^2\ln y_2  \right)}{y_1 - y_2},\\
    F_3(y_1)      &= 2y_1- 2 y_1 \ln y_1.
  \end{aligned}
\end{equation}
The constraints are applied as for the fermion mediators.

\subsection{Unitarity}\label{sSec:CaseIIUnitarity}
The unitarity constraints coming from scattering two Higgs bosons to two $\phi$'s can be obtained for the fermion mediators, albeit they are much easier to account for because of the absence of polarization for scalars. Including in $a_0$ a factor of $1/\sqrt{2}$ for identical incoming particles, the constraints are simply given by
\begin{equation}\label{eq:CaseIIUnitarityI}
  \begin{aligned}
    &\text{Case II:}\\
    &\text{max}\left(\left|\text{Re}\left(a_0^{\text{eig}}\right)\right|\right) = \frac{1}{16\sqrt{2}\pi}         \left[\sum_{i,j} \left|\sum_r \lambda^r \hat{d}^{nr}_{22ij}  \right|^2 \right]^{\frac{1}{2}} < \frac{1}{2},\\
    &\text{Case III:}\\
    & \text{max}\left(\left|\text{Re}\left(a_0^{\text{eig}}\right)\right|\right) = \frac{1}{16\sqrt{2}\pi}         \left[\sum_{i,j} \left|\sum_r \lambda^r \hat{d}^{pnr}_{22ij} \right|^2 \right]^{\frac{1}{2}} < \frac{1}{2},\\
    &\text{Case IV:}\\
    & \text{max}\left(\left|\text{Re}\left(a_0^{\text{eig}}\right)\right|\right) = \frac{|\lambda|}{16\sqrt{2}\pi} \left[\sum_{i,j} \left|                 \hat{d}^{pn}_{22ij}  \right|^2 \right]^{\frac{1}{2}} < \frac{1}{2}.\\
  \end{aligned}
\end{equation}
Note that, for case IV, since there is a single coefficient, the constraint can be simplified to
\begin{equation}\label{eq:CaseIIUnitarityII}
  |\lambda| < 8 \pi \sqrt{\frac{6}{p}}.
\end{equation}
No unitarity bound can be generally applied on $\mu$ for case I. Furthermore, a bound can be set on $Q'e'$ by adapting the results of Ref.~\cite{Hally:2012pu}. This gives
\begin{equation}\label{eq:Unitarityepscalar}
    |Q'e'| < \frac{\sqrt{4\pi}}{q^{1/4}},
\end{equation}
where $q= n + p$ for cases I, III and IV and $q = n$ for case~II.

\subsection{Results}\label{sSec:CaseIIResults}
Having introduced the relevant constraints, we now discuss the limits on $\text{BR}(h \to A A')$ for the scalar models. The sampling is performed as for the fermion mediators. Note that it is sometimes possible for a mediator to obtain a negative mass square. This would result in the breaking of $U(1)'$ and potentially the electromagnetic group. Such points are discarded because they do not correspond to the desired massless dark photon scenario and are excluded if they break electromagnetism.

\begin{figure*}[t!]
\begin{minipage}[b]{.97\textwidth}
\begin{center}
 \captionsetup[subfigure]{justification=centerlast}
 \begin{subfigure}{0.49\textwidth}
    \centering
    \caption{}
    \includegraphics[width=1\textwidth]{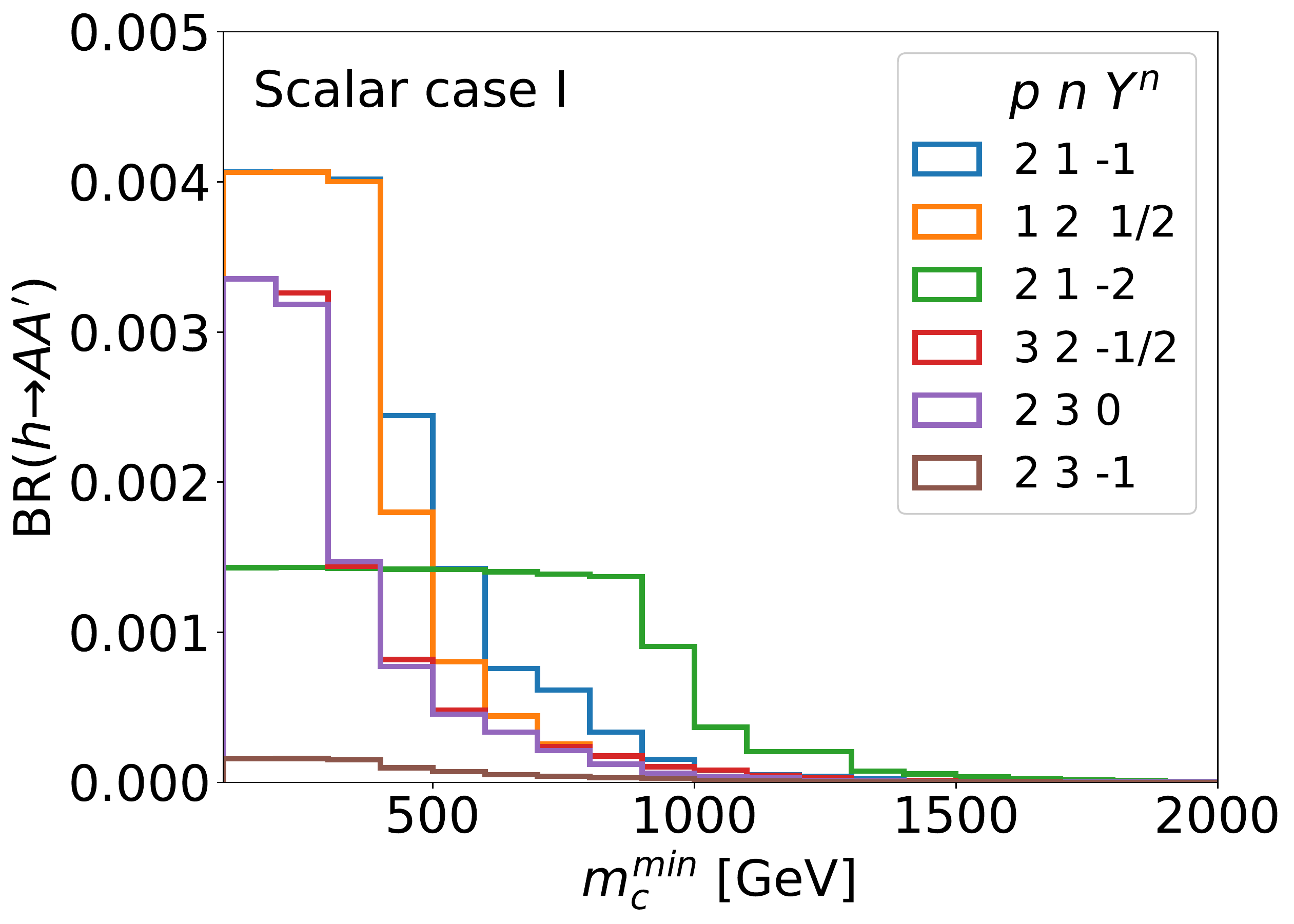}
    \label{fig:Scalar_I}
 \end{subfigure}
 \captionsetup[subfigure]{justification=centerlast}
 \begin{subfigure}{0.49\textwidth}
    \centering
    \caption{}
    \includegraphics[width=1\textwidth]{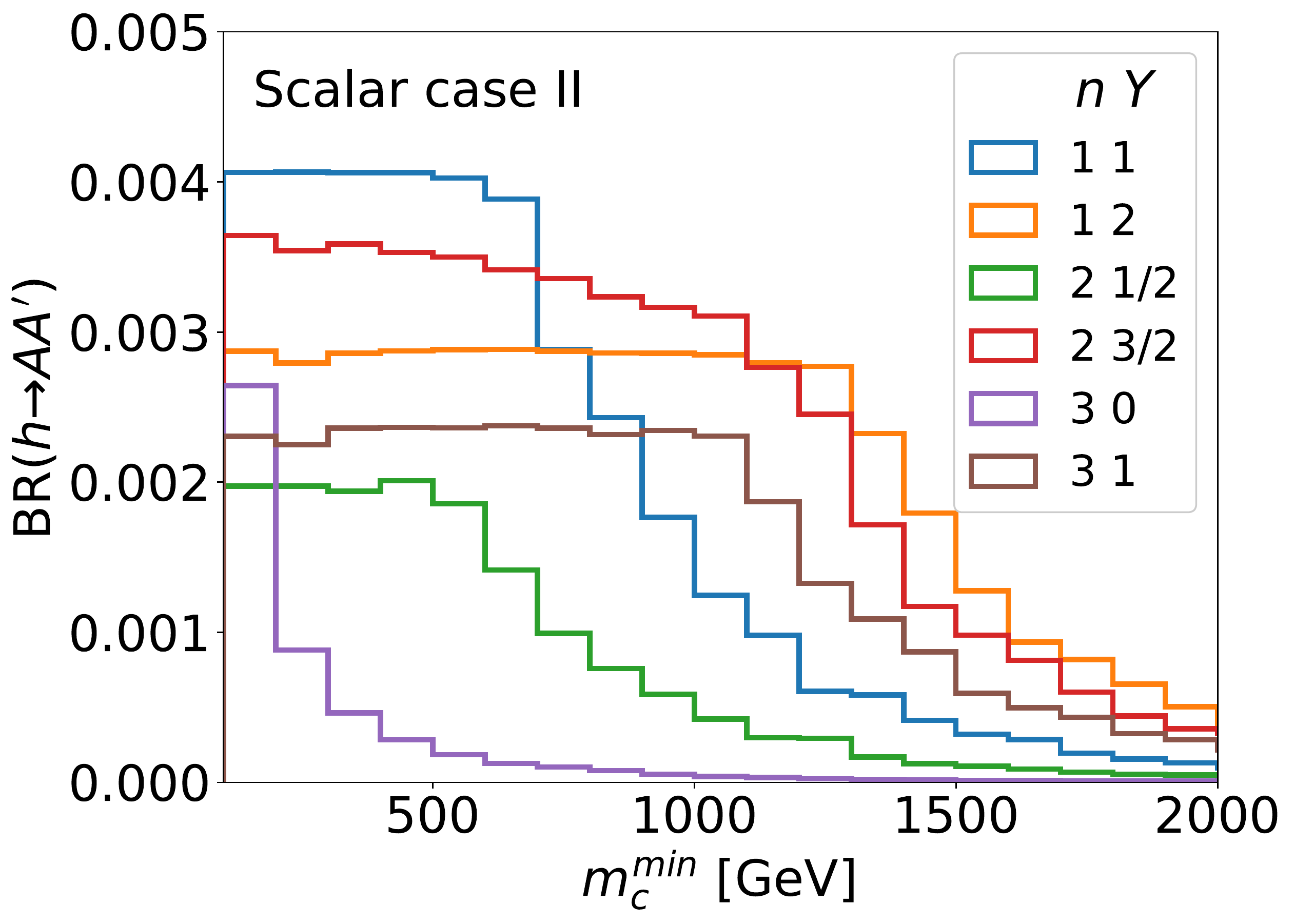}
    \label{fig:Scalar_II}
 \end{subfigure}
 \captionsetup[subfigure]{justification=centerlast}
 \begin{subfigure}{0.49\textwidth}
    \centering
    \caption{}
    \includegraphics[width=1\textwidth]{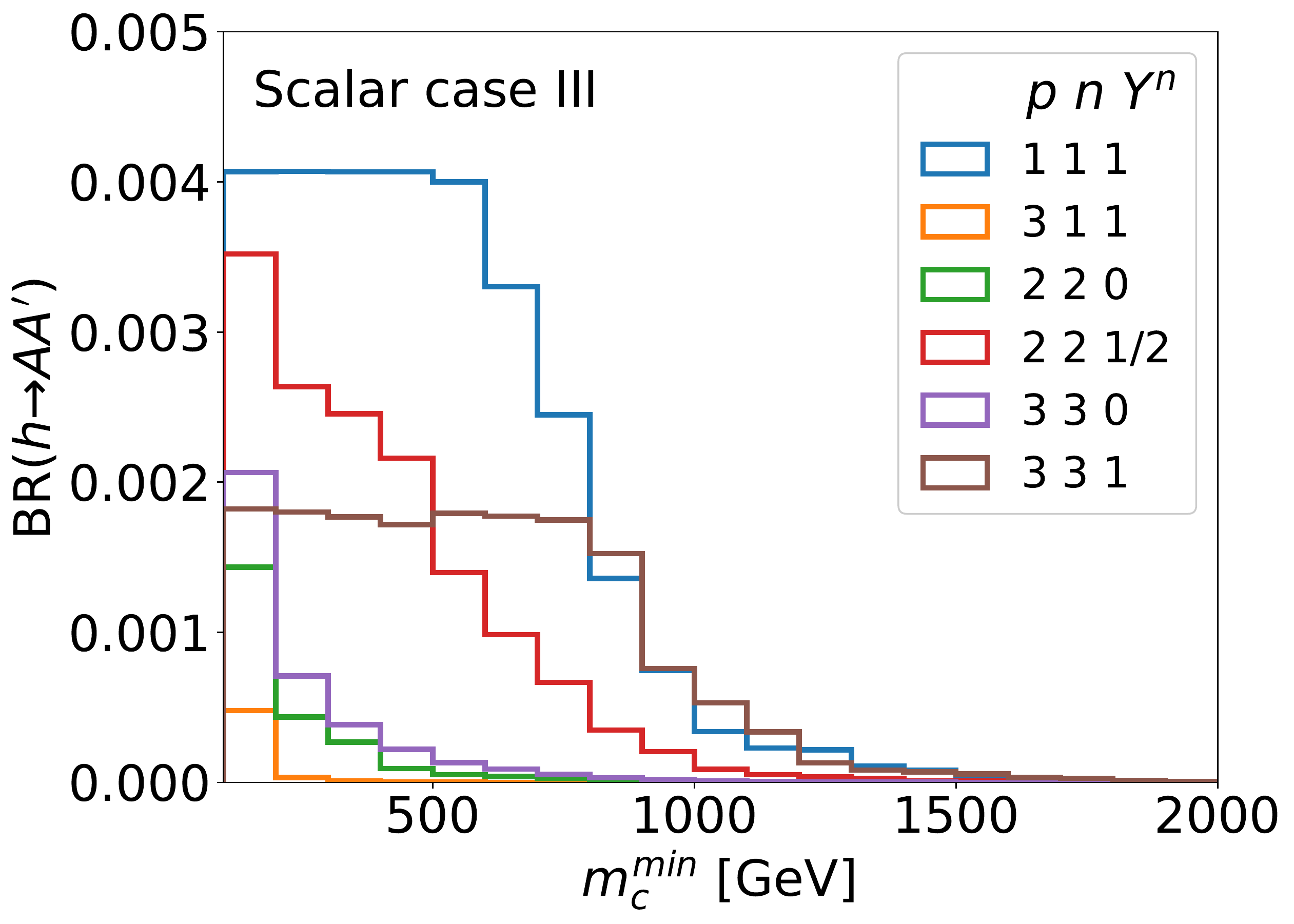}
    \label{fig:Scalar_III}
 \end{subfigure}
 \captionsetup[subfigure]{justification=centerlast}
 \begin{subfigure}{0.49\textwidth}
    \centering
    \caption{}
    \includegraphics[width=1\textwidth]{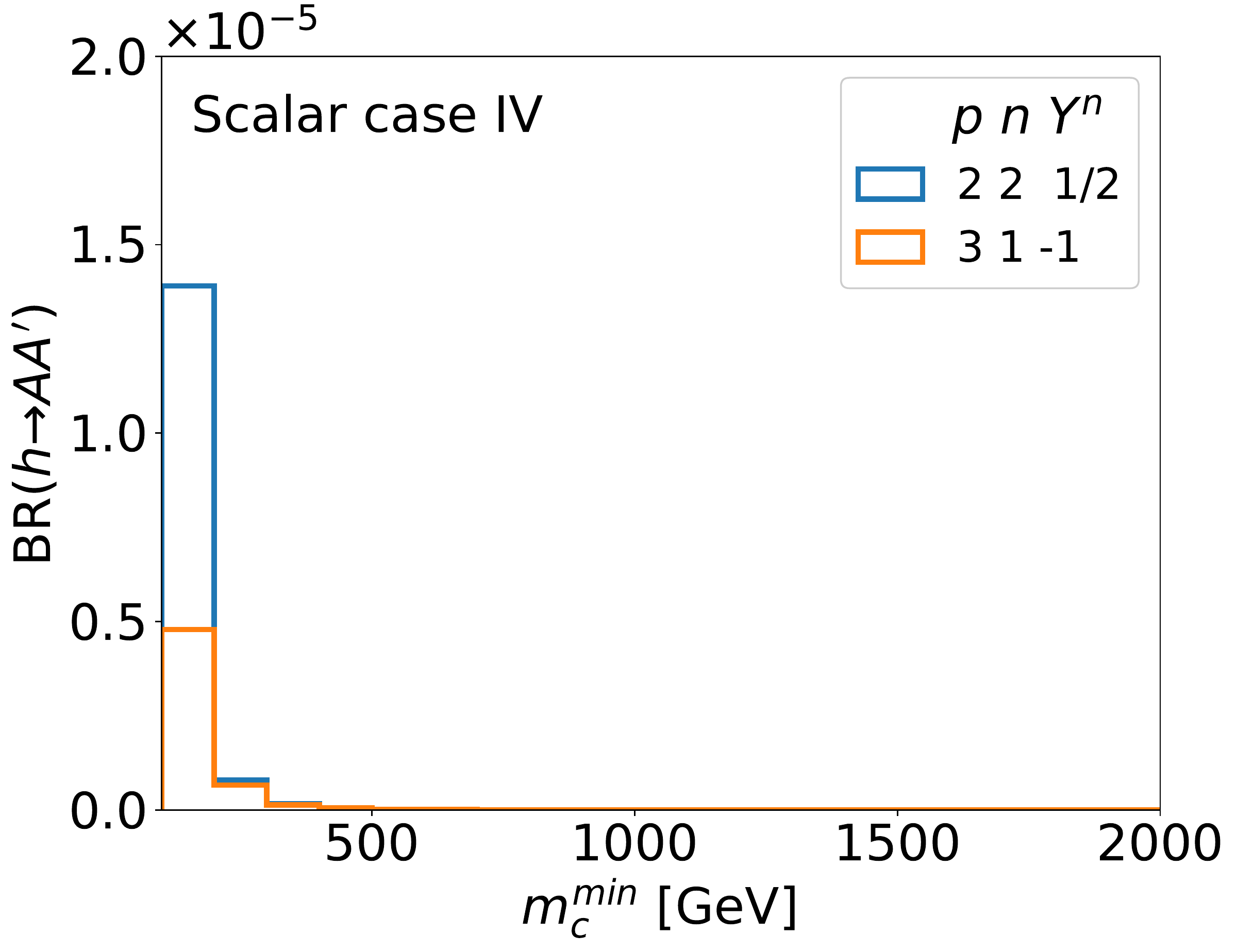}
    \label{fig:Scalar_IV}
 \end{subfigure}
\caption{Upper bounds on $\text{BR}(h \to A A')$ for different examples of scalar mediators. The plots do not go below 100~GeV, as LEP bounds prohibit such masses \cite{LEP1, LEP2}. Taken and expanded from Ref.~\cite{PhysRevLett.130.141801}. (a) Scalar case I, (b) scalar case II, (c) scalar case III, and (d) scalar case IV.}
\label{fig:ScalarPlot}
\end{center}
\end{minipage}
\end{figure*}

The plots in Fig.~\ref{fig:ScalarPlot} show the upper bounds on $\text{BR}(h \to A A')$ for models I-IV for different combinations of their gauge quantum numbers. Several comments are in order.

\begin{itemize}
  \item As can be seen, $\text{BR}(h \to A A')$ again often exhibits a plateau in the low mass regime. In the cases considered, it is still at best $0.4\%$. However, the plateau is sometimes lower because of multiple particles having similar masses. Since all mediators have identical $U(1)'$ charges but not all of them are always electrically charged, this tends to lead to a larger Higgs branching ratio to invisible particles for a given $\text{BR}(h \to A A')$.
  \item Some models are subject to far stronger constraints because of the oblique parameters or unitarity.
  \item Case I is similar to the fermion mediators and the limits are qualitatively similar.
  \item Case II can potentially avoid contributions to the oblique parameters by an appropriate choice of couplings. In practice, this is by only including $\delta_{ab}\delta_{cd}H^{a\dagger} H^b \phi^{c\dagger} \phi^d$. This scenario, however, leads to particles of similar masses that all contribute to the invisible decay of the Higgs boson. It is easy to verify that degenerate masses would have
\begin{equation}\label{eq:CaseIIIPlateaus}
  \begin{aligned}
    & \text{BR}(h \to AA') \approx \frac{1}{1 + \frac{n^2 - 1}{12 Y^2}}\\
    & \hspace{0.5cm} \times\sqrt{\text{BR}(h\to A'A') \text{BR}(h\to AA)} \left|\frac{\Delta \text{BR}(h \to AA)}{\text{BR}(h\to AA)}\right|.
  \end{aligned}
\end{equation}
which is $\lesssim 0.4\%$. However, Eq.~\eqref{eq:CaseIIIPlateaus} results in a limit of 0 when $Y = 0$. In this case, obtaining a large $\text{BR}(h \to A A')$ requires breaking the mass degeneracy which reintroduces the limits from the oblique parameters. For a general $Y$, it is not trivial whether degenerate or non-degenerate masses lead to a larger $\text{BR}(h \to A A')$.
  \item Case III is mostly similar to case II. The only difference is when $p \neq n$.  In this case, the bounds from the oblique parameters cannot be evaded and $\text{BR}(h \to A A')$ is strongly constrained.
  \item Case IV is especially constrained because it leads to a negative contribution to the $T$ parameter. This is why we only include two examples.
\end{itemize}

\section{Conclusion}\label{Sec:Conclusion}
Many collider searches have been performed in the hope of observing the Higgs boson decaying to a photon and a dark photon. For this decay to have a branching ratio realistically observable at the LHC, there must exist new mediators that communicate between the SM particles and the dark photon. In this paper, we have studied the constraints from the Higgs signal strengths, oblique parameters, EDM of the electron, and unitarity on a large set of mediator models. The models are only asked to satisfy a very minimal set of requirements. We have found that for these models, $\text{BR}(h \to A A')$ is generally constrained to be below 0.4\%, which is far lower than the current collider limit of 1.8\%. Furthermore, obtaining this 0.4\% requires relatively light charged mediators that would have somehow evaded existing searches. For some models, the bounds are even more stringent.

In addition to these constraints, a large $\text{BR}(h \to A A')$ imposes several requirements on models that might not be subjectively very pleasing. First, it requires some couplings to be very large. In hindsight, this is unsurprising. The top loop contributes relatively little to the $h \to AA$ decay width compared to $W$ loops and $\text{BR}(h \to AA)$ is still only of $\mathcal{O}(0.1\%)$. This is despite the fact that the top has a Yukawa coupling with the Higgs of $\sim 1$ and a mass of only $\sim 173$~GeV. Therefore, obtaining a large $\text{BR}(h \to A A')$ requires large couplings between the Higgs and the mediators and also a large dark electric charge $e'$. In the case of Yukawa couplings, they can be of an order of a few or larger. Second, this also leads to the presence of a Landau pole for $U(1)'$ at low energies, sometimes as low as the TeV scale.

Because of both these model requirements and the experimental constraints, we believe it would be very challenging to observe the Higgs boson decaying to a photon and a massless dark photon at the LHC.

Nonetheless, there could, in principle, be a few ways to obtain an observable $\text{BR}(h \to AA')$ by breaking some of our assumptions. Namely, it could be possible to have multiple mediators with different electric charges, $U(1)'$ charges or couplings with the Higgs boson. In this case, it might be possible to have destructive interference in channels that are particularly constrained, like $h \to AA$ and $h \to A'A'$, but constructive one in $h \to AA'$. This could, in principle, alleviate the signal strength constraints. However, reaching an unexcluded $\text{BR}(h \to A A')$ as high as 1.8\% would surely require a considerable amount of fine-tuning. Whether a channel that requires considerable fine-tuning to simply be potentially observable is worth dedicated experimental searches is certainly debatable.

\acknowledgments
This work was supported by the National Science and Technology Council under Grant No. NSTC-111-2112-M-002-018- MY3, the Ministry of Education (Higher Education Sprout Project NTU-112L104022), and the National Center for Theoretical Sciences of Taiwan.

\bibliography{biblio}

\providecommand{\href}[2]{#2}\begingroup\raggedright\begin{thebibliography}{10}

\bibitem{Holdom:1985ag}
B.~Holdom, ``{Two U(1)'s and Epsilon Charge Shifts},''
  \href{http://dx.doi.org/10.1016/0370-2693(86)91377-8}{{\em Phys. Lett. B}
  {\bfseries 166} (1986) 196--198}.

\bibitem{Spergel:1999mh}
D.~N. Spergel and P.~J. Steinhardt, ``{Observational evidence for
  selfinteracting cold dark matter},''
  \href{http://dx.doi.org/10.1103/PhysRevLett.84.3760}{{\em Phys. Rev. Lett.}
  {\bfseries 84} (2000) 3760--3763},
  \href{http://arxiv.org/abs/astro-ph/9909386}{{\ttfamily
  arXiv:astro-ph/9909386}}.

\bibitem{XENON:2020rca}
{ XENON} Collaboration, E.~Aprile {\em et~al.}, ``{Excess electronic recoil
  events in XENON1T},''
  \href{http://dx.doi.org/10.1103/PhysRevD.102.072004}{{\em Phys. Rev. D}
  {\bfseries 102} no.~7, (2020) 072004},
  \href{http://arxiv.org/abs/2006.09721}{{\ttfamily arXiv:2006.09721
  [hep-ex]}}.

\bibitem{Chiang:2020hgb}
C.-W. Chiang and B.-Q. Lu, ``{Evidence of a simple dark sector from XENON1T
  excess},'' \href{http://dx.doi.org/10.1103/PhysRevD.102.123006}{{\em Phys.
  Rev. D} {\bfseries 102} no.~12, (2020) 123006},
  \href{http://arxiv.org/abs/2007.06401}{{\ttfamily arXiv:2007.06401
  [hep-ph]}}.

\bibitem{Dobrescu:2004wz}
B.~A. Dobrescu, ``{Massless gauge bosons other than the photon},''
  \href{http://dx.doi.org/10.1103/PhysRevLett.94.151802}{{\em Phys. Rev. Lett.}
  {\bfseries 94} (2005) 151802},
  \href{http://arxiv.org/abs/hep-ph/0411004}{{\ttfamily arXiv:hep-ph/0411004}}.

\bibitem{Fradette:2014sza}
A.~Fradette, M.~Pospelov, J.~Pradler, and A.~Ritz, ``{Cosmological Constraints
  on Very Dark Photons},''
  \href{http://dx.doi.org/10.1103/PhysRevD.90.035022}{{\em Phys. Rev. D}
  {\bfseries 90} no.~3, (2014) 035022},
  \href{http://arxiv.org/abs/1407.0993}{{\ttfamily arXiv:1407.0993 [hep-ph]}}.

\bibitem{Lu:2021uec}
B.-Q. Lu and C.-W. Chiang, ``{Probing dark gauge boson with observations from
  neutron stars},'' \href{http://dx.doi.org/10.1103/PhysRevD.105.123017}{{\em
  Phys. Rev. D} {\bfseries 105} no.~12, (2022) 123017},
  \href{http://arxiv.org/abs/2107.07692}{{\ttfamily arXiv:2107.07692
  [hep-ph]}}.

\bibitem{Gabrielli:2014oya}
E.~Gabrielli, M.~Heikinheimo, B.~Mele, and M.~Raidal, ``{Dark photons and
  resonant monophoton signatures in Higgs boson decays at the LHC},''
  \href{http://dx.doi.org/10.1103/PhysRevD.90.055032}{{\em Phys. Rev. D}
  {\bfseries 90} no.~5, (2014) 055032},
  \href{http://arxiv.org/abs/1405.5196}{{\ttfamily arXiv:1405.5196 [hep-ph]}}.

\bibitem{Biswas:2015sha}
S.~Biswas, E.~Gabrielli, M.~Heikinheimo, and B.~Mele, ``{Higgs-boson production
  in association with a dark photon in e$^{+}$e$^{-}$ collisions},''
  \href{http://dx.doi.org/10.1007/JHEP06(2015)102}{{\em JHEP} {\bfseries 06}
  (2015) 102}, \href{http://arxiv.org/abs/1503.05836}{{\ttfamily
  arXiv:1503.05836 [hep-ph]}}.

\bibitem{Biswas:2016jsh}
S.~Biswas, E.~Gabrielli, M.~Heikinheimo, and B.~Mele, ``{Dark-Photon searches
  via Higgs-boson production at the LHC},''
  \href{http://dx.doi.org/10.1103/PhysRevD.93.093011}{{\em Phys. Rev. D}
  {\bfseries 93} no.~9, (2016) 093011},
  \href{http://arxiv.org/abs/1603.01377}{{\ttfamily arXiv:1603.01377
  [hep-ph]}}.

\bibitem{Biswas:2017lyg}
S.~Biswas, E.~Gabrielli, M.~Heikinheimo, and B.~Mele, ``{Dark-photon searches
  via $ZH$ production at $e^+e^-$ colliders},''
  \href{http://dx.doi.org/10.1103/PhysRevD.96.055012}{{\em Phys. Rev. D}
  {\bfseries 96} no.~5, (2017) 055012},
  \href{http://arxiv.org/abs/1703.00402}{{\ttfamily arXiv:1703.00402
  [hep-ph]}}.

\bibitem{Biswas:2017anm}
S.~Biswas, E.~Gabrielli, M.~Heikinheimo, and B.~Mele, ``{Searching for massless
  Dark Photons at the LHC via Higgs boson production},''
  \href{http://dx.doi.org/10.22323/1.314.0315}{{\em PoS} {\bfseries
  EPS-HEP2017} (2017) 315}.

\bibitem{CMS:2019ajt}
{ CMS} Collaboration, A.~M. Sirunyan {\em et~al.}, ``{Search for dark photons
  in decays of Higgs bosons produced in association with Z bosons in
  proton-proton collisions at $ \sqrt{s} $ = 13 TeV},''
  \href{http://dx.doi.org/10.1007/JHEP10(2019)139}{{\em JHEP} {\bfseries 10}
  (2019) 139}, \href{http://arxiv.org/abs/1908.02699}{{\ttfamily
  arXiv:1908.02699 [hep-ex]}}.

\bibitem{CMS:2020krr}
{ CMS} Collaboration, A.~M. Sirunyan {\em et~al.}, ``{Search for dark photons
  in Higgs boson production via vector boson fusion in proton-proton collisions
  at $ \sqrt{s} $ = 13 TeV},''
  \href{http://dx.doi.org/10.1007/JHEP03(2021)011}{{\em JHEP} {\bfseries 03}
  (2021) 011}, \href{http://arxiv.org/abs/2009.14009}{{\ttfamily
  arXiv:2009.14009 [hep-ex]}}.

\bibitem{ATLAS:2021pdg}
{ ATLAS} Collaboration, G.~Aad {\em et~al.}, ``{Observation of electroweak
  production of two jets in association with an isolated photon and missing
  transverse momentum, and search for a Higgs boson decaying into invisible
  particles at 13~$\text {TeV}$ with the ATLAS detector},''
  \href{http://dx.doi.org/10.1140/epjc/s10052-021-09878-z}{{\em Eur. Phys. J.
  C} {\bfseries 82} no.~2, (2022) 105},
  \href{http://arxiv.org/abs/2109.00925}{{\ttfamily arXiv:2109.00925
  [hep-ex]}}.

\bibitem{ATLAS:2022xlo}
{ ATLAS} Collaboration, ``{Search for dark photons from Higgs boson decays via
  $ZH$ production with a photon plus missing transverse momentum signature from
  $pp$ collisions at $\sqrt{s}$ = 13 TeV with the ATLAS detector},''
  \href{http://arxiv.org/abs/2212.09649}{{\ttfamily arXiv:2212.09649
  [hep-ex]}}.

\bibitem{Bjorken:1988as}
J.~D. Bjorken, S.~Ecklund, W.~R. Nelson, A.~Abashian, C.~Church, B.~Lu, L.~W.
  Mo, T.~A. Nunamaker, and P.~Rassmann, ``{Search for Neutral Metastable
  Penetrating Particles Produced in the SLAC Beam Dump},''
  \href{http://dx.doi.org/10.1103/PhysRevD.38.3375}{{\em Phys. Rev. D}
  {\bfseries 38} (1988) 3375}.

\bibitem{Curtin:2014cca}
D.~Curtin, R.~Essig, S.~Gori, and J.~Shelton, ``{Illuminating Dark Photons with
  High-Energy Colliders},''
  \href{http://dx.doi.org/10.1007/JHEP02(2015)157}{{\em JHEP} {\bfseries 02}
  (2015) 157}, \href{http://arxiv.org/abs/1412.0018}{{\ttfamily arXiv:1412.0018
  [hep-ph]}}.

\bibitem{Chang:2016ntp}
J.~H. Chang, R.~Essig, and S.~D. McDermott, ``{Revisiting Supernova 1987A
  Constraints on Dark Photons},''
  \href{http://dx.doi.org/10.1007/JHEP01(2017)107}{{\em JHEP} {\bfseries 01}
  (2017) 107}, \href{http://arxiv.org/abs/1611.03864}{{\ttfamily
  arXiv:1611.03864 [hep-ph]}}.

\bibitem{Fox:2018ldq}
P.~J. Fox, I.~Low, and Y.~Zhang, ``{Top-philic $Z'$ forces at the LHC},''
  \href{http://dx.doi.org/10.1007/JHEP03(2018)074}{{\em JHEP} {\bfseries 03}
  (2018) 074}, \href{http://arxiv.org/abs/1801.03505}{{\ttfamily
  arXiv:1801.03505 [hep-ph]}}.

\bibitem{Parker:2018vye}
R.~H. Parker, C.~Yu, W.~Zhong, B.~Estey, and H.~M\"uller, ``{Measurement of the
  fine-structure constant as a test of the Standard Model},''
  \href{http://dx.doi.org/10.1126/science.aap7706}{{\em Science} {\bfseries
  360} (2018) 191}, \href{http://arxiv.org/abs/1812.04130}{{\ttfamily
  arXiv:1812.04130 [physics.atom-ph]}}.

\bibitem{Pan:2018dmu}
J.-X. Pan, M.~He, X.-G. He, and G.~Li, ``{Scrutinizing a massless dark photon:
  basis independence},''
  \href{http://dx.doi.org/10.1016/j.nuclphysb.2020.114968}{{\em Nucl. Phys. B}
  {\bfseries 953} (2020) 114968},
  \href{http://arxiv.org/abs/1807.11363}{{\ttfamily arXiv:1807.11363
  [hep-ph]}}.

\bibitem{Fabbrichesi:2020wbt}
M.~Fabbrichesi, E.~Gabrielli, and G.~Lanfranchi, ``{The Dark Photon},''
  \href{http://arxiv.org/abs/2005.01515}{{\ttfamily arXiv:2005.01515
  [hep-ph]}}.

\bibitem{PhysRevLett.130.141801}
H.~Beauchesne and C.-W. Chiang, ``Is the decay of the higgs boson to a photon
  and a dark photon currently observable at the lhc?,''
  \href{http://dx.doi.org/10.1103/PhysRevLett.130.141801}{{\em Phys. Rev.
  Lett.} {\bfseries 130} (Apr, 2023) 141801}.
  \url{https://link.aps.org/doi/10.1103/PhysRevLett.130.141801}.

\bibitem{Biswas:2022tcw}
S.~Biswas, E.~Gabrielli, and B.~Mele, ``{Dark Photon Searches via Higgs Boson
  Production at the LHC and Beyond},''
  \href{http://dx.doi.org/10.3390/sym14081522}{{\em Symmetry} {\bfseries 14}
  no.~8, (2022) 1522}, \href{http://arxiv.org/abs/2206.05297}{{\ttfamily
  arXiv:2206.05297 [hep-ph]}}.

\bibitem{Passarino:1978jh}
G.~Passarino and M.~J.~G. Veltman, ``{One Loop Corrections for e+ e-
  Annihilation Into mu+ mu- in the Weinberg Model},''
  \href{http://dx.doi.org/10.1016/0550-3213(79)90234-7}{{\em Nucl. Phys. B}
  {\bfseries 160} (1979) 151--207}.

\bibitem{Patel:2015tea}
H.~H. Patel, ``{Package-X: A Mathematica package for the analytic calculation
  of one-loop integrals},''
  \href{http://dx.doi.org/10.1016/j.cpc.2015.08.017}{{\em Comput. Phys.
  Commun.} {\bfseries 197} (2015) 276--290},
  \href{http://arxiv.org/abs/1503.01469}{{\ttfamily arXiv:1503.01469
  [hep-ph]}}.

\bibitem{LEP1}
{ LEPSUSYWG, ALEPH, DELPHI, L3 and OPAL collaboration} Collaboration,
  ``{Combined lep chargino results, up to 208 gev for large m0}.''
\newblock
  \url{http://lepsusy.web.cern.ch/lepsusy/www/inos_moriond01/charginos_pub.html}.

\bibitem{LEP2}
{ LEPSUSYWG, ALEPH, DELPHI, L3 and OPAL collaboration} Collaboration,
  ``{Combined LEP Chargino Results, up to 208 GeV for low DM}.''
\newblock
  \url{http://lepsusy.web.cern.ch/lepsusy/www/inoslowdmsummer02/charginolowdm_pub.html}.

\bibitem{Heinemeyer:2013tqa}
{ LHC Higgs Cross Section Working Group} Collaboration, J.~R. Andersen {\em
  et~al.}, ``{Handbook of LHC Higgs Cross Sections: 3. Higgs Properties},''
  \href{http://arxiv.org/abs/1307.1347}{{\ttfamily arXiv:1307.1347 [hep-ph]}}.

\bibitem{ATLAS:2023tkt}
{ ATLAS} Collaboration, ``{Combination of searches for invisible decays of the
  Higgs boson using 139 fb$^{-1}$ of proton-proton collision data at $\sqrt{s}
  = 13$ TeV collected with the ATLAS experiment},''
  \href{http://arxiv.org/abs/2301.10731}{{\ttfamily arXiv:2301.10731
  [hep-ex]}}.

\bibitem{CMS-PAS-HIG-19-005}
{ CMS Collaboration} Collaboration, ``{Combined Higgs boson production and
  decay measurements with up to 137 fb-1 of proton-proton collision data at
  sqrts = 13 TeV},'' tech. rep., CERN, Geneva, 2020.
\newblock \url{http://cds.cern.ch/record/2706103}.

\bibitem{ATLAS-CONF-2021-053}
{ ATLAS Collaboration} Collaboration, ``{Combined measurements of Higgs boson
  production and decay using up to $139$ fb$^{-1}$ of proton-proton collision
  data at $\sqrt{s}= 13$ TeV collected with the ATLAS experiment},'' tech.
  rep., CERN, Geneva, Nov, 2021.
\newblock \url{http://cds.cern.ch/record/2789544}.

\bibitem{Nakai:2016atk}
Y.~Nakai and M.~Reece, ``{Electric Dipole Moments in Natural Supersymmetry},''
  \href{http://dx.doi.org/10.1007/JHEP08(2017)031}{{\em JHEP} {\bfseries 08}
  (2017) 031}, \href{http://arxiv.org/abs/1612.08090}{{\ttfamily
  arXiv:1612.08090 [hep-ph]}}.

\bibitem{Chang:2005ac}
D.~Chang, W.-F. Chang, and W.-Y. Keung, ``{Electric dipole moment in the split
  supersymmetry models},''
  \href{http://dx.doi.org/10.1103/PhysRevD.71.076006}{{\em Phys. Rev. D}
  {\bfseries 71} (2005) 076006},
  \href{http://arxiv.org/abs/hep-ph/0503055}{{\ttfamily arXiv:hep-ph/0503055}}.

\bibitem{Roussy:2022cmp}
T.~S. Roussy {\em et~al.}, ``{A new bound on the electron's electric dipole
  moment},'' \href{http://arxiv.org/abs/2212.11841}{{\ttfamily arXiv:2212.11841
  [physics.atom-ph]}}.

\bibitem{ACME:2018yjb}
{ ACME} Collaboration, V.~Andreev {\em et~al.}, ``{Improved limit on the
  electric dipole moment of the electron},''
  \href{http://dx.doi.org/10.1038/s41586-018-0599-8}{{\em Nature} {\bfseries
  562} no.~7727, (2018) 355--360}.

\bibitem{Peskin:1990zt}
M.~E. Peskin and T.~Takeuchi, ``{A New constraint on a strongly interacting
  Higgs sector},'' \href{http://dx.doi.org/10.1103/PhysRevLett.65.964}{{\em
  Phys. Rev. Lett.} {\bfseries 65} (1990) 964--967}.

\bibitem{Anastasiou:2009rv}
C.~Anastasiou, E.~Furlan, and J.~Santiago, ``{Realistic Composite Higgs
  Models},'' \href{http://dx.doi.org/10.1103/PhysRevD.79.075003}{{\em Phys.
  Rev. D} {\bfseries 79} (2009) 075003},
  \href{http://arxiv.org/abs/0901.2117}{{\ttfamily arXiv:0901.2117 [hep-ph]}}.

\bibitem{Lavoura:1992np}
L.~Lavoura and J.~P. Silva, ``{The Oblique corrections from vector - like
  singlet and doublet quarks},''
  \href{http://dx.doi.org/10.1103/PhysRevD.47.2046}{{\em Phys. Rev. D}
  {\bfseries 47} (1993) 2046--2057}.

\bibitem{Chen:2003fm}
M.-C. Chen and S.~Dawson, ``{One loop radiative corrections to the rho
  parameter in the littlest Higgs model},''
  \href{http://dx.doi.org/10.1103/PhysRevD.70.015003}{{\em Phys. Rev. D}
  {\bfseries 70} (2004) 015003},
  \href{http://arxiv.org/abs/hep-ph/0311032}{{\ttfamily arXiv:hep-ph/0311032}}.

\bibitem{Carena:2007ua}
M.~Carena, E.~Ponton, J.~Santiago, and C.~E.~M. Wagner, ``{Electroweak
  constraints on warped models with custodial symmetry},''
  \href{http://dx.doi.org/10.1103/PhysRevD.76.035006}{{\em Phys. Rev. D}
  {\bfseries 76} (2007) 035006},
  \href{http://arxiv.org/abs/hep-ph/0701055}{{\ttfamily arXiv:hep-ph/0701055}}.

\bibitem{Chen:2017hak}
C.-Y. Chen, S.~Dawson, and E.~Furlan, ``{Vectorlike fermions and Higgs
  effective field theory revisited},''
  \href{http://dx.doi.org/10.1103/PhysRevD.96.015006}{{\em Phys. Rev. D}
  {\bfseries 96} no.~1, (2017) 015006},
  \href{http://arxiv.org/abs/1703.06134}{{\ttfamily arXiv:1703.06134
  [hep-ph]}}.

\bibitem{Cheung:2020vqm}
K.~Cheung, W.-Y. Keung, C.-T. Lu, and P.-Y. Tseng, ``{Vector-like Quark
  Interpretation for the CKM Unitarity Violation, Excess in Higgs Signal
  Strength, and Bottom Quark Forward-Backward Asymmetry},''
  \href{http://dx.doi.org/10.1007/JHEP05(2020)117}{{\em JHEP} {\bfseries 05}
  (2020) 117}, \href{http://arxiv.org/abs/2001.02853}{{\ttfamily
  arXiv:2001.02853 [hep-ph]}}.

\bibitem{Zyla:2020zbs}
{ Particle Data Group} Collaboration, P.~Zyla {\em et~al.}, ``{Review of
  Particle Physics},'' \href{http://dx.doi.org/10.1093/ptep/ptaa104}{{\em PTEP}
  {\bfseries 2020} no.~8, (2020) 083C01}. and 2021 update.

\bibitem{Hally:2012pu}
K.~Hally, H.~E. Logan, and T.~Pilkington, ``{Constraints on large scalar
  multiplets from perturbative unitarity},''
  \href{http://dx.doi.org/10.1103/PhysRevD.85.095017}{{\em Phys. Rev. D}
  {\bfseries 85} (2012) 095017},
  \href{http://arxiv.org/abs/1202.5073}{{\ttfamily arXiv:1202.5073 [hep-ph]}}.

\end{thebibliography}\endgroup
\bibliographystyle{utphys}

\end{document}